*Article*

# A Comprehensive Framework for Analyzing IoT Platforms: A Smart City Industrial Experience

**Mahdi Fahmideh [1,*], Jun Yan [1], Jun Shen [1], Davoud Mougouei [1], Yanlong Zhai [2], and Aakash Ahmad [3]**

[1] University of Wollongong, Australia; jyan@uow.edu.au (J.Y.); jshen@uow.edu.au (J.S.); davoud@uow.edu.au (D.M.)
[2] Beijing Institute of Technology Beijing, China; ylzhai@bit.edu.cn
[3] University of Ha'il, Saudi Arabia; a.abbasi@uoh.edu.sa
* Correspondence: Mahdi@uow.edu.au

**Abstract:** The compliance of IoT platforms to quality is paramount to achieve users' satisfaction. Currently, we do not have a comprehensive set of guidelines to appraise and select the most suitable IoT platform architectures that meet relevant criteria. This paper is a tentative response to this critical knowledge gap where we adopted the design science research approach to develop a novel evaluation framework. Our research, on the one hand, stimulates an unbiased competition among IoT platform providers and, on the other hand, establishes a solid foundation for IoT platform consumers to make informed decisions in this multiplicity. The application of the framework is illustrated in example scenarios. Moreover, lessons learned from applying design science research are shared.

**Keywords:** design science research; IoT platform; evaluation framework; architecture evaluation; decision making; smart city

## 1. Introduction

Many IT-based organizations have shown their interest in embarking on IoT platforms to implement and integrate the next generation of data-intensive Internet-based applications. IoT platforms provide the required infrastructure and core services for connectivity and managing a large amount of data, dealing with a large variety of smart objects, and improved scalability. The recent surveys [1,2] pinpoint that the exponential growth and the variety of existing platforms offer the flexibility of choice, but, on the flip side, they imply that, prior to a subscription, users need to ensure the quality of platforms.

From the perspective of an IoT platform user or consumer, who views an IoT platform as a black box, it becomes a challenging exercise to select one or more platforms that fulfill business requirements as each platform offers features that are common with other platforms and distinct features at different levels and costs. That is, reliable decision making depends on an understanding of pertinent quality factors contributing to a quality IoT platform as well as a rigorous way to qualitatively/quantitatively measure these factors. This understanding mitigates from a cursory selection and unforeseen contingencies which may result in the poor quality and maintenance overhead cost of applications that are deployed on a platform. Furthermore, practitioners may need to compare and rank IoT platforms based on their specific set of requirements with different priorities.

On the other hand, from the perspective of an IoT platform provider or provider who has a white-box view of an IoT platform, it is important to ensure the quality of a designed IoT platform architecture before jumping prematurely into the discussion on enabling technologies to operationalize the platform architecture design. This assessment is needed in a high-level architecture design endeavor where there is more flexibility to tackle the potential flaws in the architecture instead of identifying and rectifying such flaws during the platform operation, which can be costly.

A critical need for both IoT platform consumers and providers is a set of checklist, guidance, and metrics enabling the suitability assessment of an IoT platform that is going to provide a backbone for other applications. As discussed in Section 2, there is a paucity of research on providing a comprehensive guidance framework for IoT platform analysis and selection. Little is known about the quality criteria for incorporation into the evaluation and selection of IoT platforms, and this knowledge gap has been largely remained unaddressed in the relevant literature [2–4].

We, UTX (abbreviation for us as the university), were approached by an IoT service provider in Australia (IoT Australia, hereafter), which collaborates with a broad coalition of partners from local, state, and federal government to provide people-centric IoT-based solutions. IoT Australia asked us to propose an evaluation framework to use to effectively appraise the quality of its IoT platform reference architecture called ROSE. ROSE was a base architecture model to design domain-specific IoT-based architecture at New South Wales (NSW) in Australia. Our project brief session with the CEO (chief executive officer) and senior research consultant of IoT Australia highlighted that their development team wanted to operationalize ROSE for a municipal council at NSW. Nevertheless, they were not sure if ROSE was complete and adhered to quality features. Therefore, the IoT Australia asked us to address the following important practical research objective: *To introduce a criteria-based framework to support the evaluation and comparison of IoT platform architectures*. We compiled a list of commonly agreed criteria derived from the IoT literature together with measuring guidance to provide a basis for an in-depth evaluation of IoT platforms. The application of the framework in assessing the ROSE architecture and ranking a representative sample of existing IoT platforms are also presented.

We applied the Design Science Research (DSR) approach [5] to achieve our research objective. The DSR relies on the theory-based development of a design artefact for real-world problems and an interplay between design and use of the artefact [6]. Adopting this approach enabled us to make a link between theories (academics) and practice (professionals) effectively and to produce theoretically supported and empirically validated prescriptive an IT artifact. To the best of our knowledge, this paper is the first attempt to address the deficiency of the literature on both evaluation frameworks and application of the DSR in an industry project, all while in the context of IoT. Our findings benefit both researchers and practitioners contemplating DSR for designing quality-aware IoT-based applications.

This article is structured as follows: Section 2 provides background on IoT platform architecture, evaluation frameworks, and the problem of IoT platform evaluation and selection. Section 3 explains our DRS approach adopted to develop and validate the proposed framework. Section 4 presents the full description of the proposed framework criteria. The application of the framework is discussed in two case scenarios of the ROSE framework and ranking a representative set of IoT platforms in Section 5. This is followed by a discussion on research implications, limitations, and surveying related studies in Section 6. Finally, this article concludes in Section 7.

## 2. Research Background

Typically, an IoT-based system constitutes a few key building elements, namely, people, applications, infrastructures, things, and platforms, as shown in Figure 1. An IoT platform in such a harness is a key element. It is defined as an integrated middleware environment relying on enabling technologies such as cloud computing, big data analytics platforms, and cyber-physical systems to provide necessary infrastructure and core services to enable the development and integration of applications for different domains, such as smart cities and manufacturing [1].

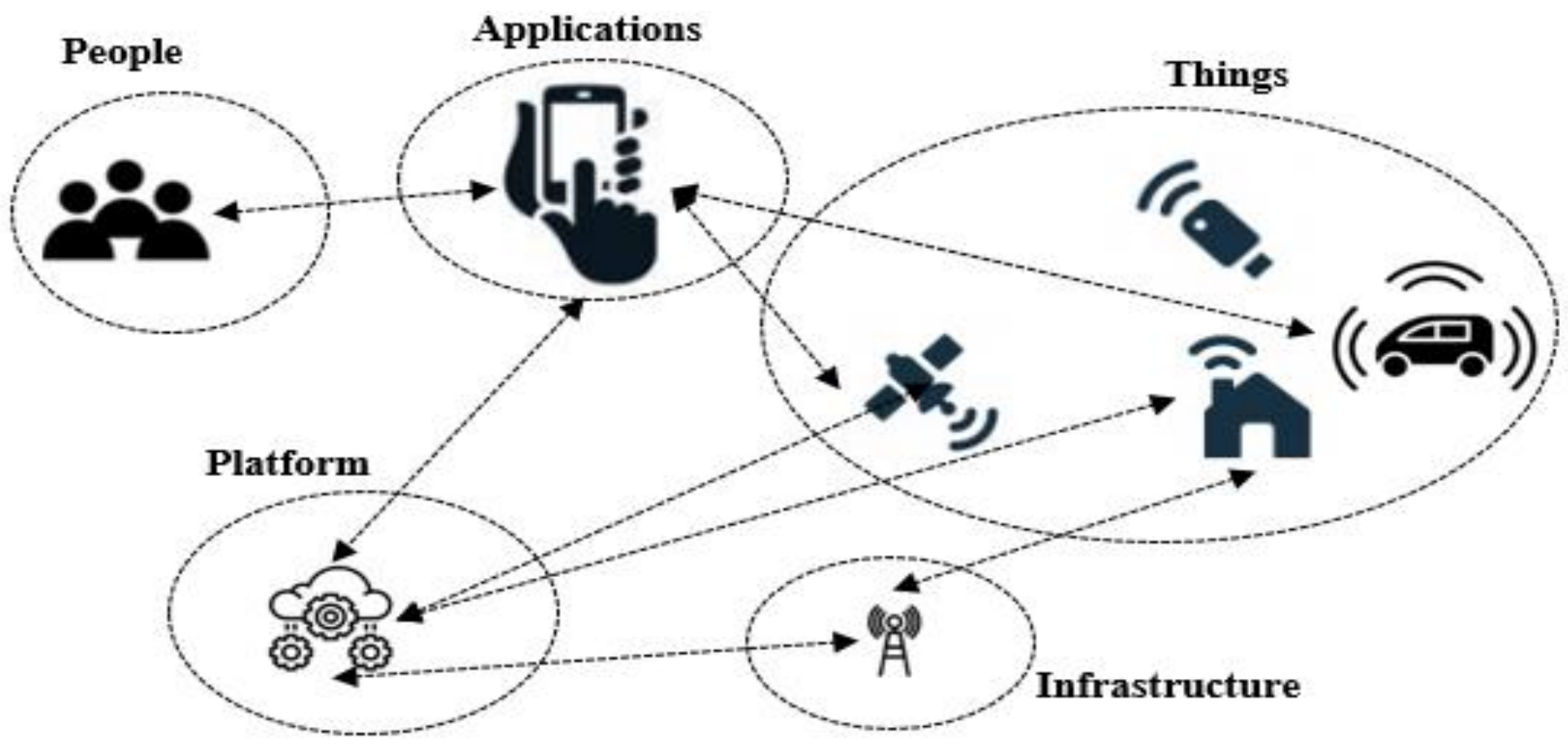

**Figure 1.** Key elements of an IoT-based system.

We view an IoT platform architecture from the perspective of software architecture that is defined in conventional software engineering, i.e., breaking the platform down into its high-level interconnected structure such as layers and software/hardware components [7]. Several architectural representations with a different recognition of layers exist in IoT literature; though when collectively viewed, they have intersections. For instance, Komninos's defines three architecture layers, namely, information storage, application, and user interface [8]. From a people–system interaction perspective, Al-Hader et al. extend the layers to five including infrastructure, database, building management systems, smart interface, and smart city [9]. Furthermore, Luca's suggested that an IoT-based architecture consists of two layers, i.e., knowledge processors and semantic information brokers. In line with these definitions, the architecture of an IoT platform may constitute the following layers:

- The user interface layer provides interactivity between users and platform services via providing dashboards, reports, message boards, 3D spaces, and 2D maps so that users can get access services based on the subject of interest and send queries.
- The application layer places software systems such as mission-critical business systems, e.g., ERP (enterprise resource planning), mobile applications, business analytics applications/reports, back-end services, and management and monitoring applications. The layer enables users to receive data from smart objects, perform processing algorithms over the data, and send the results back to users or other objects.
- The service layer provides IoT backend services and application programming interfaces (APIs) required for end-to-end application development that are positioned in the application layer. E-government services, traffic control, citizen security, water conservation, social services, and environmental protection are examples of services. Adequately designed, the layer reduces the cost and effort for implementing applications for the application layer.
- The data layer keeps data from various data sources floating across the operational environment such as real-time data from sensors, geospatial data, historical data, and social media.
- The infrastructure layer provides hardware for data processing, storing, computing, and interconnectivity among data centers, servers, and smart objects via different network communication protocols.

Increasing the size and complexity of software systems has triggered some of the most important issues that have increased interest in proposing quality-driven architecture evaluation frameworks. They are a means to lead system architects to examine if a proposed architecture would result in desired quality and to identify any potential design flows in advance [7,10]. While the idea of using architecture evaluation frameworks for assessing the quality of systems has been well recognized in the information system design and software engineering disciplines, it has received less attention in the IoT community. Analogically, it is evident that an IoT platform architecture can be viewed as a type of software architecture [11,12]. Taking into account this analogy and the insights we gave during meetings with the IoT Australia's development team, we believed the evaluation frameworks are a suitable tool for incorporation into the early stage of IoT platform architecture design and assessment. The need for having such evaluation frameworks is important, as it was evident by our industry partner IoT Australia, if the conformance to a standard quality is enforced. One of the areas, which is particularly highlighted in this research, is the critical assessment and selection among a number of platforms in light of user's quality expectations. On the platform consumer side, it has become difficult to decide which platform is the right choice among a large list of alternatives in the market and what should be the basis for the selection before the subscription to a particular platform. On the platform provider side, it is important to identify design flaws at the high-level design and tackle them in advance where changes to the architecture are less costly. The identification of such flaws at the early stage of architecture design endeavor can be facilitated if a suitable yardstick is available.

## 3. Research Approach

Aiming at the development of a purposeful artefact, i.e., the evaluation framework, to be sensed and used by IoT Australia for appraising its own IoT platform architecture, i.e., ROSE, in municipal council at NSW, it was beneficial to follow a particular and defined research paradigm suitable for this genre of research. We conducted DSR approach [5] phases, commonly used by researchers in software engineering and information systems disciplines. It includes iterative phases of design and validation. We specialized DSR phases in the context of our research as depicted in Figure 2. Our DSR journey began with the call made by the IoT Australia stating the lack of an evaluation framework to improve the reliability of a design IoT platform architecture for the municipal city council at NSW.

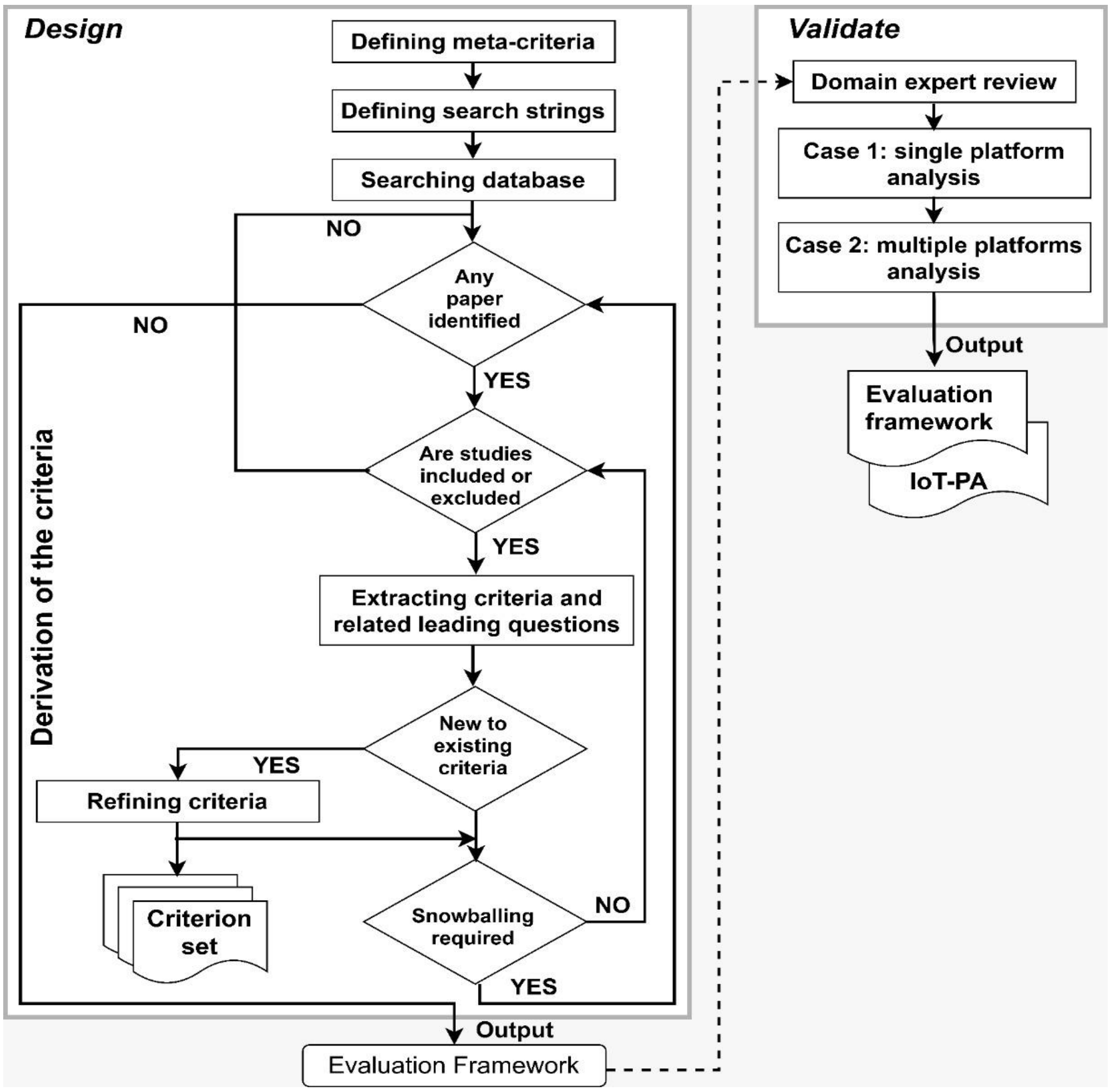

**Figure 2.** Design science phases conducted in this research.

### 3.1. Design

Kernel theories, approaches, techniques, or concepts provide a foundation for context-specific artefact design endeavors. In this regard, we leveraged software architecture evaluation [7], a systematic literature review (SLR) [13], and ongoing models of the IoT Australia, and relevant knowledge about the inner operating environment of the IoT Australia during the framework design phase. The derivation of the criterion set was performed in two steps as follows.

Step 1. Meta-criteria definition. To achieve a fair set of evaluation criteria, the first step was to specify meta-criteria (MC) (the criteria for evaluation other criteria) that would be expected to be satisfied by a criterion set. We borrowed MC from the existing software engineering evaluation frameworks [14,15], presented in Table 1 with slight refinements for this research. They were anchored during the compilation and checking the suitability of our proposed evaluation criteria.

**Table 1.** MCs were anchored during the development, validation, and evaluation of criteria.

| | |
|---|---|
| MC1 (Simplicity) | A criterion should be clear and easy to understand by IoT architecture assessors. |
| MC2 (Preciseness) | A criterion should be detailed, unambiguous, and measurable to highlight the similarities and differences between different instances of an IoT architecture. |
| MC3 (Minimum overlapping) | Criteria should be distinct and have a minimum dependency or likeness to each other. |
| MC4 (Soundness) | Criteria should make sense for IoT experts and have a semantic link to IoT domain. The soundness of criteria can be assessed if it is mandatory or only desirable. |
| MC5 (Generality) | Criteria should be sufficiently generic and be applicable to evaluate all IoT architecture regardless of underlying implementation technologies. |
| MC6 (Consistency) | Criteria should not have a conflict with each other and contrast. |
| MC7 (Balance) | Criteria should consider as many aspects of an IoT architecture design. |
| MC8 (Measurability) | A criterion should provide a means as possible to measure it either or both in qualitative or quantitative ways. It should be noted that our framework inclines to be on qualitative measurements as quantities were not realistic at the early stage of IoT architecture design. |
| MC9 (Comprehensiveness) | Though it is unrealistic, criteria should be inclusive as much as possible and cover all important facets relevant to IoT architecture design with respect to the evaluation objectives. |

Step 2. Criterion set derivation. To initiate the development of the criteria, we leveraged the literature pertinent to IoT. Given the proliferation of publications in this field of research, we believed SLR [13] would be an appropriate way to synthesize the set of criteria. The SLR is a well-recognized method to identify, analyze, and interpret all related studies with respect to a topic of interest. We determined platform, smart city, evaluation, and requirements as the main terms to define the search queries and extended them with alternative synonyms. The search terms were combined to derive alternative search queries using logical operations AND and OR as shown in Table 2.

**Table 2.** Search queries (SQ) used during literature review.

| | |
|---|---|
| SQ1 | "Platform", "Architecture" OR "Reference" OR "Reference architecture" OR "Layer" OR "Reference Model" OR "Layered architecture" OR "Stack" OR "Concept" OR "Model", OR "Platform" AND [SQ2 OR SQ3 OR SQ4] |
| SQ2 | "Smart city" OR "IoT" OR "Smart IoT" |
| SQ3 | "Evaluation" OR "Assessment" OR "Framework" OR "Methodology" OR "Approach" |
| SQ4 | "Requirements" OR "Challenges" OR "Concerns" OR "Issues" OR "Needs" OR "Quality" |

We used the search queries against the scientific digital libraries such as Google Scholar, IEEE Explore, ACM Digital Library, Springer Link, and Science Direct to identify papers that were as follows: (i) explicitly proposing an IoT platform, IoT architecture, conceptual model, surveys, or challenges related to the design and development of IoT platform architecture design, and (ii) published between 2008 and May 2019. The starting date of 2008 was based on the Cisco's estimation about the emergence of IoT. Moreover, our literature review continued till 2019, i.e., our tentative project time frame. Before 2008, the search strings did not yield any relevant results. We excluded papers that were in languages other than English from our review. Moreover, among the multiple papers from an author, we selected the most recent and completed one for the review. Appendix A shows 63 identified papers with their demographic information as the source for defining the evaluation criteria. We read the abstract, introduction, and conclusion sections of each paper to identify texts, quotes, recommendations, or challenges that could refer to a meaningful criterion or a question that could lead to a qualitative assessment question. Other than that, it is important to realize that assessing the identified platforms, as listed in [16], is not the objective of this research. Rather, the evaluation served as the evidence to support the importance of the proposed criterion set in our evaluation framework. Indeed, as the platforms are continuously being evolved, the evaluation needs to be updated.

Moreover, the criteria were compiled through an iterative refinement process as discussed in [1]. That is, the derivation of the criterion set has been based on (i) the top-down approach where we took into account general literature on conventional software architecture evaluation to get an insight of important criteria and (ii) the bottom-up approach where different existing IoT-specific studies as shown in Appendix A were analyzed to specialize the criterion set with the IoT context and required leading questions.

MCs, which were set in step 1, were interleaved during the criterion set compilation. In adherence to MC9 (comprehensiveness), we used the snowballing technique [17] into the review process to include as many relevant papers as possible. That is, the papers that were cited in the reference and related work sections of the under review paper were fed into the next round of the literature review iteration to identify new criteria. For MC4 (soundness), we applied a simple rule. That is, the frequent occurrence of a criterion grounded in the selected studies was an indicator of the meaningfulness and importance of that criterion. With regard to MC5 (generality), we included the criteria that were sufficiently generic to consider in the variety of IoT architecture designs and filtered out those relevant to a specific IoT platform vendor or application domain. For MC7 (balance), we classified the criteria into two groups of functional and non-functional based on the previously published IoT literature surveys[18–20].

### *3.2. Domain Expert Review*

We obtained qualitative comments from domain experts in IoT platform architecture design regarding our framework's adherence to the meta-criteria defined in the design phase. This step gave us an opportunity to refine the framework via modifying, adding, and removing the criteria. Two experts participated and individually reviewed the documentation of the framework: (i) one academic with experience in enterprise architecture design (external to this project) as denoted by E1 and (ii) the senior research consultant of the IoT Australia (from our project partner team) as denoted by E2.

The feedback from the experts was analyzed, and necessary refinements were applied to the framework, resulting in the next version presented in the next section. An area of concern raised by E1 was the lack of orthogonality between the criteria and the IoT architecture layers, i.e., it is important to design criteria/leading questions that examine all the layers. This concern was in line with the meta-criterion preciseness. For example, E1 mentioned the leading questions for the criterion security should be defined in a way that it informs an assessor of checking the security from the user interface layer down toward the physical layer. Based on the literature sources, we defined new questions per criterion in relation to each layer (Appendix B). In Section 4, we discuss how each criterion is related to the layers as a point of interest for the architecture assessment. Furthermore, E1 suggested providing an online version of the framework for simply the execution of an assessment endeavor. With respect to the meta-criterion comprehensiveness, E2 suggested adding a criterion called data sharing to the evaluation framework. We believed that the data sharing is subsumed under the criterion security and adding it as a new criterion to the framework could lead to a criteria redundancy. Thus, it was not considered as an individual criterion in our framework. At the end of this evaluation step, the refined version of the framework was obtained, which is presented in Tables 3 and 4 and detailed in Section 4. The second and third steps of validation to demonstrate the applicability of the framework in practice are presented in Section 5.

**Table 3.** Criteria related to assessing IoT platform functions (UL: user interface layer, AL: application layer, SL: service layer, DL: data layer, PL: physical layer, √: related, ×: Not related).

| Criterion | Leading Evaluation Question(s) for Measuring Criterion | Layers | | | | |
|---|---|---|---|---|---|---|
| | | UL | AL | SL | DL | PL |
| Resource discovery | Does the platform provide the mechanisms (e.g., service discovery protocols, electronic product codes (EPC), and ubiquitous codes (uCode)) to dynamically and automatically identify new resources (e.g., devices, sensors, actuators, services, and applications that connect to the smart city network) at any time? | × | √ | √ | × | √ |
| Data accumulation | Does the platform provide mechanisms for continuous data collection from various objects for processing across all layers? | × | √ | √ | × | √ |
| Data cleaning | Does the platform provide mechanisms for identifying and correcting inaccurate or incomplete data before storing them into data storage? | × | √ | √ | × | √ |
| Data storing | Does the platform use proper data storage mechanisms for storing various structured and unstructured data collected from the environment? | × | × | × | √ | × |
| Data analysis | Does the platform implement mechanisms for identifying useful knowledge from data and predicting the environment's behavior? | × | × | √ | × | × |
| Query processing | Does the platform provide a language so that user is enabled to send their query over data sources connected to the platform for data retrieval? | × | √ | √ | × | √ |
| Generating meta-data | Does the platform provide mechanisms to store data and meta-data associated with objects, users, applications, and SLAs for better management and reflection on platform performance? | × | √ | √ | × | √ |
| Data visualization | Does the platform provide proper visualizations such as diagrams, tables, charts, and icons for presenting data analysis results to users? | √ | × | × | × | × |
| Continuous monitoring | Does the platform define mechanisms for internally monitoring and assessing its components to detect violations from quality factors across all the layers? Does the platform have mechanisms for externally monitoring the environment to detect unexpected behaviors? | × | √ | √ | √ | √ |
| Service composition | Does the platform provide APIs to enable users to compose individual services offering by the platforms or other applications with respect to their needs? | × | √ | √ | √ | √ |
| Event processing | Does the platform define mechanisms for responding to internal events happening inside the platform components? Does the platform define mechanisms for responding to external events happening in the environment? | × | √ | √ | √ | √ |

**Table 4.** Criteria related to assessing IoT platform functions (UL: user interface layer, AL: application layer, SL: service layer, DL: data layer, PL: physical layer, √: related, ×: Not related).

| Criterion | Leading Evaluation Question(s) for Measuring Criterion | Layers | | | | |
|---|---|---|---|---|---|---|
| | | UL | AL | SL | DL | PL |
| Security | Does the architecture implement mechanisms to keep the security of data and objects across the platform layers? | √ | √ | √ | √ | √ |
| | Does the platform limit the data acquisition from users' device and smart objects to keep security and privacy of users? | √ | √ | √ | √ | √ |
| | Does the platform enable a user to define customized smart objects discoverability at different levels? These levels can be, for example, public where an object can be discovered to all users, private where an object is discoverable by the object owner, and friends where the object is discoverable by a friend key provided by the virtual object. | × | × | × | × | √ |
| | Does the platform provide mechanisms and techniques for a safe exchange of data and interactions among technical objects across layers? | √ | √ | √ | √ | √ |
| | Does the platform define an access control list on routers, packet filters, firewalls, and network-based intrusion detection systems across layers? | × | × | × | × | √ |
| | Does the platform define authentication and authorization mechanisms? | √ | √ | √ | √ | √ |
| | Does the platform define encryption/decryption mechanisms across layers? | √ | √ | √ | √ | √ |
| | Does the platform define functional decomposition (separating functional components) to avoid propagating security issues to other platform components? | × | √ | √ | × | × |
| Privacy | Does the platform provide mechanisms such as encryption/decryption for protecting the personal data of users such as citizens or organizations using the platform or data collected from them by the platform? Does the platform have proper authentication and authorization mechanisms to protect unauthorized access? | √ | √ | √ | √ | √ |
| Interoperability | Does the platform provide integration mechanisms, protocols, or middleware such as HTTP, MQTT (message queuing telemetry transport which is a broker-based publishing/subscribing), and AMQP (advanced message queuing protocol) for integrating applications? | × | √ | × | × | × |
| | Does the platform use mechanisms to exchange and process heterogeneous events coming from multiple sources at different levels of the system? | × | √ | √ | √ | √ |
| | Does the platform have mechanisms, e.g., ontologies, to unify interoperability points to collect, process, and generate data from/to diverse data sources, legacy devices, and objects? | × | × | × | √ | × |
| | Does the platform provide mechanisms such as web service, virtual object, representational State Transfer/ReST to interact with other IoT platforms in order to call/use their services/data? | × | × | √ | × | × |
| | Does the platform provide mechanisms to identify heterogeneous resources in a network such as sensor networks, IP networks? | × | × | × | × | √ |
| | Does the platform provide the necessary interfaces to facilitate invoking its services? | × | × | √ | × | × |
| | Does the platform provide open APIs to avoid vendor lock-in issues? | × | √ | √ | √ | √ |
| Reusability | Does the platform have templates of objects/services to be used for other scenarios different from the default? | × | √ | √ | × | × |
| | Does the platform have the necessary interfaces to hide service incompatibilities and enable the reuse of calls by its services? | × | √ | √ | × | × |
| | Does the platform allow modifying and reusing its core functions/services for applications? | × | √ | √ | × | × |
| | Does the platform apply architecture design principles such as loose coupling, modularity, or layering to facilitate reuse? | × | √ | √ | × | × |
| | Does the platform have simple and small services instead of coarse grain and composite services? | × | √ | √ | × | × |
| | Can the platform operate across multiple connectivity protocols? | × | √ | √ | √ | √ |
| Reliability | Does the platform have mechanisms to be resilient to failures in order to address service availability in the case of a fault in its components? | × | √ | √ | √ | √ |
| | Does the platform ensure users of the success rate and is being error-free in the provided services? | × | √ | √ | √ | √ |
| | Does the platform's architecture define redundant servers or replication mechanisms to avoid the single point of failure? | × | √ | √ | × | √ |
| | Does the platform have mechanisms such as anomaly detections for reliability and accuracy of data collected from a variety of sources which can be noisy? | × | × | × | × | × |
| | Does the platform have mechanisms to ensure that new smart objects and components joining the network are compatible together to avoid the unreliability of the platform? | × | × | × | × | √ |
| Recoverability | Does the platform provide mechanisms and techniques to restore to a state in which it can continue performing its functions after the occurrence of a hazard/failure? | × | √ | √ | √ | √ |
| Availability | Does the platform provide mechanisms for the continuous guarantee of obtaining, storing, processing, and providing data and services to users independently of the state of underlying infrastructure? | × | √ | √ | √ | √ |
| | Does the platform provide component/data clustering or replication/distribution mechanisms to increase availability? | × | √ | √ | √ | √ |
| | Does the platform provide redundant storage arrays e.g., RAID (redundant array of independent disks) for the data layer? | × | × | × | √ | × |
| | Does the platform provide redundant physical links for the network layer? | × | × | × | × | √ |
| | Does the platform provide redundant nodes, e.g., routers allowing service discovery, for PL? | × | × | × | × | √ |
| Extensibility | Does the platform apply mechanisms or techniques such as defining extension points in components' function, interfaces, templates, or modularization principle to enable adding new objects such as sensors, datatypes, and urban contexts to itself to support new functionalities? | × | √ | √ | × | × |
| Adaptability | Does the platform define mechanisms such as machine learning algorithms, rules, or policies to support the dynamic transformation from the current configuration to a new configuration or in the case of occurring disruptive events? | × | √ | √ | × | × |
| | Does the platform have mechanisms for incorporating upcoming functional, data, and objects to accommodate new platform configuration toward continuous improvement? | × | √ | √ | × | × |
| Scalability | Does the platform include mechanisms to effective processing algorithms to guarantee the best possible throughput of services offering to users with minimum cost? | × | √ | √ | × | × |
| Performance | Does the platform include mechanisms to effective processing algorithms to guarantee the best possible throughput of services offering to users with minimum cost? | × | √ | √ | × | × |
| Usability | Does the platform provide suitable and effective user interfaces such as dashboards, reports, message boards, 3D spaces, 2D maps for interaction with and presenting information to users? | √ | × | × | × | × |
| Configurability | Does the platform have mechanisms for its users to change the behavior or settings of components to suit new changes? | × | √ | √ | × | √ |
| | Does the platform have mechanisms for the configuration of layers (vertical configuration) such as smart objects, services, network protocols, devices, switches, permissions, and user access? | × | × | × | × | √ |
| | Does the platform allow users to configure support of run-time configuration of its components including adding or removing components? | × | √ | √ | × | × |
| | Does the platform support mechanisms for platform configuration in relation to other external platforms (horizontal configuration)? | × | × | × | × | √ |
| | Does the platform support the run-time configuration of its components including adding or removing components? | × | √ | √ | × | × |
| Mobility | Does the platform provide mechanisms for mobility protocols in the support of continuous user movement to different locations? | × | × | × | × | √ |
| Efficiency | Does the platform define mechanisms to estimate energy consumption by its internal and external components? | × | √ | √ | × | √ |
| | Does the platform define mechanisms to predict user behaviors and elasticity demands to schedule energy and resource allocation through pricing models? | × | √ | √ | × | √ |
| | Does the platform have data storage related to energy logs to store a history of energy consumption by objects? | × | × | × | √ | × |
| | Does the platform log the history of its components' operations and state of objects such as changing TV channel, switching on/off light for further analysis? | × | × | × | √ | × |
| | Does the platform have data storage related to the environment to log the history of changes in the environment such as temperature, humidity, and the number of people who are typically collected by objects? | × | × | × | √ | × |
| | Does the platform implement mechanisms in smart objects to address the efficient use of resources? | × | × | × | × | √ |
| | Does the platform have mechanisms for service composition with respect to resource constraints of involved smart objects such as batty power? | × | × | √ | × | × |
| Maintainability | Does the platform simplify applying refinements, correcting faults to its components, or adding new functionalities/increments to the existing components via applying design principles such as low coupling, high cohesion, and separation of concerns? | × | √ | √ | × | × |
| | Does the platform define clear definitions of objects such as naming, icons, documentation/modelling, and interactions, as well as design principles such as modularity, less coupling, and separation of concerns, to make modifications simpler and less costly? | × | √ | √ | × | × |

### 3.3. *IoT-PA*

In addressing E1's concern, we implemented a prototype system, called IoT platform analyzer (IoT-PA). It was as an output of our design phase to facilitate working with the framework. It was developed using the standard software development life cycle and implemented based on open source technologies (e.g., PHP and MySql). As depicted in Figure 3, the IoT-PA has four functions each including interactive forms enabling an assessor to browse and evaluate either a single IoT platform or multiple cloud platforms. An assessment scenario may not be performed in one single run. Rather, users can create an assessment project and update it as the platform architecture evolves. For example, a group of platform developers can create an assessment project to evaluate the quality of their IoT platform architecture design against the selected criteria and the guidance of leading questions. If the assessment outcome is not satisfactory, the developers can save the project, address the deficiencies in the architecture, and then update the assessment to meet the quality criteria. In a multiple assessment scenario, developers may update the assessment steps as far as more information about candidate platforms are provided.

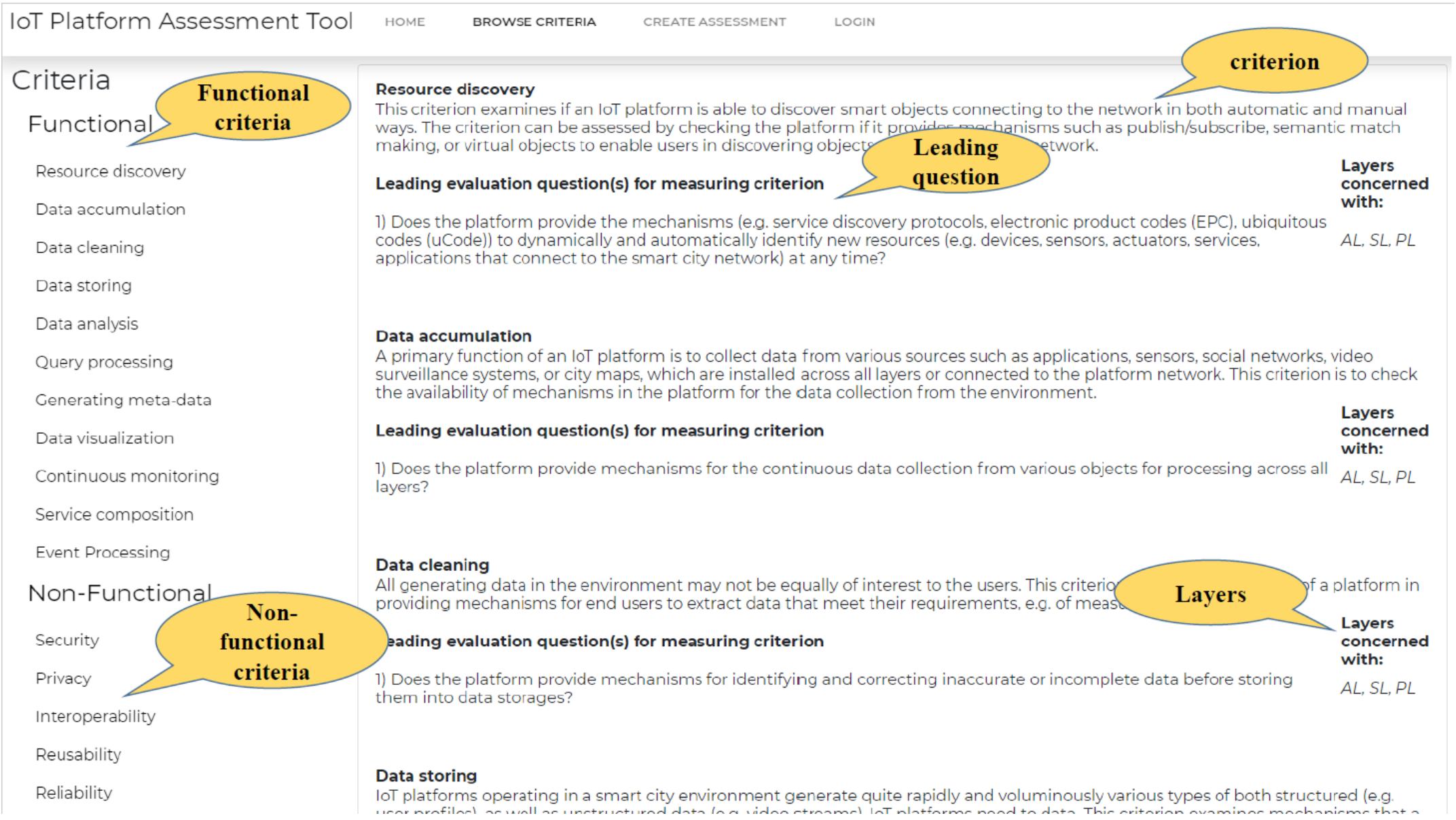

**Figure 3.** The main page of the IoT-PA, an implementation of the evaluation framework.

## 4. Evaluation Framework

The evaluation framework consists of 26 criteria classified under two dimensions: functional- and non-functional-related criteria (quality factors). In Tables 3 and 4, we used symbols √ and × to specify IoT platform architecture layers for which a criterion should be checked. The criterion set is described in the following subsections.

### 4.1. *Functional-Related Criteria*

The criteria in this dimension examines eleven key functionalities that should be supported by an IoT platform. In practice, these criteria are realized via implementing different software/hardware components and application programing interfaces (APIs). They are described in what follows.

(i) Resource discovery. This criterion examines if an IoT platform can identify objects connecting to the network in both automatic and manual ways. The criterion can be assessed by checking the platform if it provides mechanisms such as publish/subscribe, semantic match-making, or virtual objects to enable users in discovering objects or subscribe to the network.
(ii) Data accumulation. A primary function of an IoT platform is to collect data from various sources such as applications, sensors, social networks, video surveillance systems, or city maps, which are installed across all layers or connected to the platform network. This criterion is to check the availability of mechanisms in the platform for the data collection from the environment.
(iii) Data cleaning. All generating data in the environment may not be equally of interest to the users. This criterion checks the capability of a platform in providing mechanisms for users to extract data that meet their requirements, e.g., air temperature after a specific threshold value or comparison of measured values.
(iv) Data storing. IoT platforms voluminously collects various types of both structured, e.g., user profiles, as well as unstructured data, e.g., video streams implying the need to have multiple types of data storage technologies to save data. In an IoT platform, typically, two types of relational SQL-based and No-SQL data storages are used. Relational databases are a common option if atomicity, consistency, isolation, durability constraints, and support for complicated queries are required. In addition, No-SQL databases like Hadoop, CouchDB, CouchBase, MongoDB, and HBase support mechanisms for horizontal scalability, memory and distributed index, and dynamic data schema modification.
(v) Data analysis. An IoT platform is expected to provide sophisticated mechanisms for performing classification, mining, regression, and clustering over both stream and batch data. The platform capability in utilizing enabling technologies such as Apache Storm, Apache Spark, or Hadoop MapReduce is an indicator of this criterion support.
(vi) Query processing. This criterion examines the availability of languages provided by the platform enabling users to send and perform queries over data sources connected to the platform.
(vii) Generating meta-data. Meta-data, which can be associated with different hardware and software components of an IoT platform, is a means to facilitate the classification, identification, and retrieval of the data. Generating meta-data can be conducted, for instance, by users where a platform may enable annotating the components as well as platform internally during data collection and processing. This criterion considers mechanisms to handle meta-data generation in a platform.
(viii) Data visualization. Providing data visualization in a variety of form is an important function of an IoT platforms which can be examined in terms of availability of icons, diagrams, and tables to represent data analysis results and signals to users.
(ix) Continuous monitoring. This criterion examines if the platform provides mechanisms to monitor environmental fluctuations and ensure the performance of its components. To support this criterion, a platform needs to provide supports to keep the track of platform components' behavior, e.g., applications, services, smart objects, and network elements, to identify anomalies, correlations, or similar patterns of divergence. Platform's components related to the monitoring function provide this information to other services.
(x) Service composition. Different platform components may offer individual services/functionalities that can be combined to create cross-platform composite services to build complex IoT applications. This implies a need for APIs and tools for building composite services via using other existing fine-granular services.
(xi) Event processing. This criterion examines if the platform has mechanisms for representation, capturing, and quickly reacting to important internal events, e.g., system events and external ones, e.g., city events, peak-time vehicle speed, and geographic events. An event is an observable change in the state of the environment, which can be triggered by platform components, e.g., smart objects. Event processing in the platform should be handled in real-time so that users receive accurate and timely platform services.

*4.2. Non-Functional-Related Criteria*

The criteria in the second dimension of the proposed framework examines the platform support of commonly expected non-functional requirements, i.e., quality factors. It is important to realize that these criteria are orthogonal to the functional requirements. For example, the security criteria is concerned across all layers of the IoT platform. These criteria are listed in Table 4. In the following, we describe the criteria the layers for which they are concerned in an architecture assessment exercise.

(i) Security. In an operational environment where smart objects dynamically join and leave the network, robust security mechanisms should be addressed across all layers in a way that newly joining smart objects to the network still satisfy the expected platform security requirements.
(ii) Privacy. Checking this criterion is to ensure if personal user data in interactions with the platform is properly protected and restricted from unauthorized access during saving and processing.
(iii) Interoperability. The need for interoperability support is ubiquitous across all the different layers of an IoT platform to enable interaction of heterogeneous software and hardware components, e.g., smart objects, services, and applications, to ensure service delivery to users. One form of required interoperability is related to data as the platform interacts with a variety of heterogeneous data formats coming from different sources when collecting, storing, and processing data. The interoperability is also needed at the application and service layers where different platforms may need to interact with services to identify and react swiftly to events and trends. This criterion should be assessed for this layer if end users need support for service composition from heterogeneous components. The user may be interested in the platform support of exchanging and processing heterogeneous events coming from multiple sources. Moreover, the interoperability criterion should be investigated in the view of the infrastructure layer providing hardware for data storage and computation as well as physical interconnectivity among data centers, servers, networks, and devices. As presented in Table 4, in assessing platform compliance to this criterion, the user can check mechanisms and techniques such as ontologies, semantic annotation, semantic web services, linked data standards, and virtual objects in a platform, which are commonly used to address interoperability in software systems. A key principle in addressing interoperability is to define abstraction levels supported by interfaces/APIs to connect disparate technologies.
(iv) Reusability. A platform's services should be designed in such a way that users can reuse them for other purposes than the one originally offered. For example, a platform performs some fixed functions for data collection, filtering, saving, and processing. The end-user may need to modify and reuse these functions to build custom-specific applications or composite services for different applications deployed on the same platform. As shown in Table 4, for example, a platform can support the reusability criterion if it provides fine granular services (e.g., microservices) and APIs with minimum functions instead of large and coarse-grained services.
(v) Reliability. Reliability is the extent to which an IoT platform keeps its expected functions with the required precision. At the application/service layer, this criterion is meant as the capability of the platform to predict, detect, and fix unforeseen failures if they occur in its components to continue their operations. One can realize the support for this criterion through mechanisms such as replicating and distributing platform components over different servers. The criterion can be subjective in terms of the data layer. The data coming from sensors might be noisy and abnormal which affects the derived data analysis results. Due to reasons such as a low battery or broken repeater, a sensor may report a temperature that is out of the expected range or may stop reporting altogether. From the data reliability point of view, a platform needs to provide mechanism for data anomaly detection, reconciliation, semantic consistency, and normalization.

(vi) Recoverability. This criterion is to check the existence of mechanisms across different layers of a platform which restore components to a correct state to continue their operations after experiencing an unexpected incidence leading to failure.

(vii) Availability. This criterion investigates implemented mechanisms in the platform to guarantee the expected percentage of time in which the platform's services/data are in operation and accessible when required to use. The same mechanisms such as component replications, distributions, and fault resilience that are accommodated for the platform reliability can also be equally applied to improve availability. For instance, once a platform component fails to continue service provisioning, an alternative component should start working that service.

(viii) Extensibility. This criterion, which concerns with the application and service layers, refers to the extent to which a platform's services are easy to be extended to new functionalities with minimum modifications. Adherence to the extensibility criterion is required to support continuous changes in requirements of users.

(ix) Adaptability. Dynamic capability of a platform to adapt to new settings by swapping or combining its components is necessary to respond to frequent changes occur in the environment. The adaptability criterion can be supported in a platform by providing the learning ability in components to acquire knowledge from the environment and to dynamically reconfigure or optimize themselves to new situations.

(x) Scalability. Without degrading the quality criteria, the platform should be able to operate effectively when there is an increase in (a) received requests from users (b) volume of data collected, stored, and processed, and (c) number of components, e.g., smart objects interacting with the platform. If the platform cannot process requests within a threshold, more resources, e.g., adding new servers, should be allocated to the platform components to maintain the expected rate of response time. To realize the scalability, IoT platforms leverage enabling technologies such as cloud computing, data analytics, and microservices. Therefore, the evaluation of this criterion depends on the availability of mechanisms such as vertical and horizontal scalability in these technologies.

**(xi)** Performance. This criterion is to check the existence of designed mechanisms, e.g., effective processing algorithms, to guarantee the best possible response time, i.e., task completion time, and throughput, i.e., the number of processed requests per time unit.

(xii) Usability. This criterion evaluates the quality of a user's experience in interacting with the platform's services. This criterion assesses how easy and simple it is to understand, control communications, get information, and interpret presented information by the platform. The criterion can be checked at the different levels of architecture layer, for example the simplicity of smart object configuration, wherever a user can interact with one.

(xiii) Configurability. A platform should provide mechanisms for setting its functions, services, interfaces, and connecting smart objects to operate and communicate in a specific way at a certain point in time. The configurability, which may be, for example, setting sensors, congestion threshold parameters, or service priority, happens in two stages of development and execution. The former is related to configure platform layers before a user starts interacting with the platform, whilst the latter is related to a run-time reconfiguration/self-reconfiguration capability of the platform in dynamic assembling or disassembling of its components to guarantee QoS. This criterion is important to assess since deploying platform components on different infrastructures may demand different security or performance requirements.

(xiv) Mobility. In the dynamic environment of platform operation, components like smart objects change their locations and move from a platform to another platform which can be at the level of suburb, state, city, or country. This criterion is to check if the platform provides adequate support for the safe movements of smart objects without the degradation of other non-functional requirements, specifically security.

(xv) Efficiency. A platform is expected to provide effective resource management and energy consumption in its components. Internal resource management concerns with mechanisms defined in the platform to keep a record of execution of threads, track of available and utilizing resources, resource allocation algorithms, task scheduling, and performance information. In addition, at the application and physical layers, the data collection, storage, and processing functions should be resource efficient as they are typically battery powered. Furthermore, the external resource management is related to the platform support for the extensive data collection about energy consumption from houses, buildings, the number of installed devices, electric consumption, water consumption, and gas consumption.

(xvi) Maintainability. IoT platforms need to respond to frequent run-time changes in their hardware and software components. The maintainability is the ease with which its component can be modified to change or add new capabilities, correct defects, and improve other non-functional requirements. As such, platform maintenance should not be difficult and costly in fulfilment of new business requirements. Applying proper and adequate design principles such as low coupling, high cohesion, and separation concerns in platform architecture and interacting components aids maintainability. Likewise, clear and detailed documentation of the architecture, e.g., in diagrams or formal specifications, facilitates handling changes to components.

## 5. Application of Framework

The following subsections report our findings in adopting the framework in assessment of both individual and multiple platforms with respect to the criterion set. The purpose of the first case study is to examine if the framework could be a viable tool to capture important assessment aspects of ROSE. This would inform the IoT Australia of flaws of the ROSE base architecture. The second case study is to demonstrate the framework application in facilitating the comparison and ranking of IoT platforms via AHP. Whilst the former case study demonstrates the applicability of the framework to support a platform provider in a white-box based architecture assessment exercise, i.e., ROSE, the latter deals with a black-box-based assessment in favor of a platform consumer. The case studies also illustrate the adherence of our framework to the meta-criteria that have been set earlier in Section 3.1.

### *5.1. Case Study 1: Single-Platform Analysis*

The IoT Australia had decided to roll out ROSE (Figure 4), as a high-level base platform architecture for local government designing their own IoT-based smart city capacities in NSW. The ROSE architecture had been used to implement an environmental monitoring system prototype for low-cost distributed air quality, temperature, and noise monitoring in the urban settings of a local government in NSW. The system had on-board sensors for temperature, humidity, particulates, carbon monoxide, nitrogen dioxide, ozone, and noise. It had a power-optimized system that could function for 24 h in complete darkness and designed for continuous operation for up to 7 years. Since the first version of the ROSE framework had been independently designed by the IoT Australia without collaboration with the UTX, there was a concern if ROSE has been sufficiently complete and capable of capturing the important aspects of the architecture of IoT-based solutions in NSW. The CEO of the IoT Australia said:

> *"My questions revolve around the maturity of platform architectures, i.e. the ROSE project, which is positioned to be a template and a test case for IoT-based smart city* *architecture. How does the ROSE architecture stack up against developing smart city platform architecture standards? What are the pragmatic compromises to be made to accommodate real-world implementations today? .... there were no explicit architectural requirements for ROSE".*

Our framework was used to answer this question as explained in the following. This assessment gave an opportunity for contrasts between ROSE and our framework toward identifying improvement areas in both ROSE and our framework, i.e., the IoT-PA.

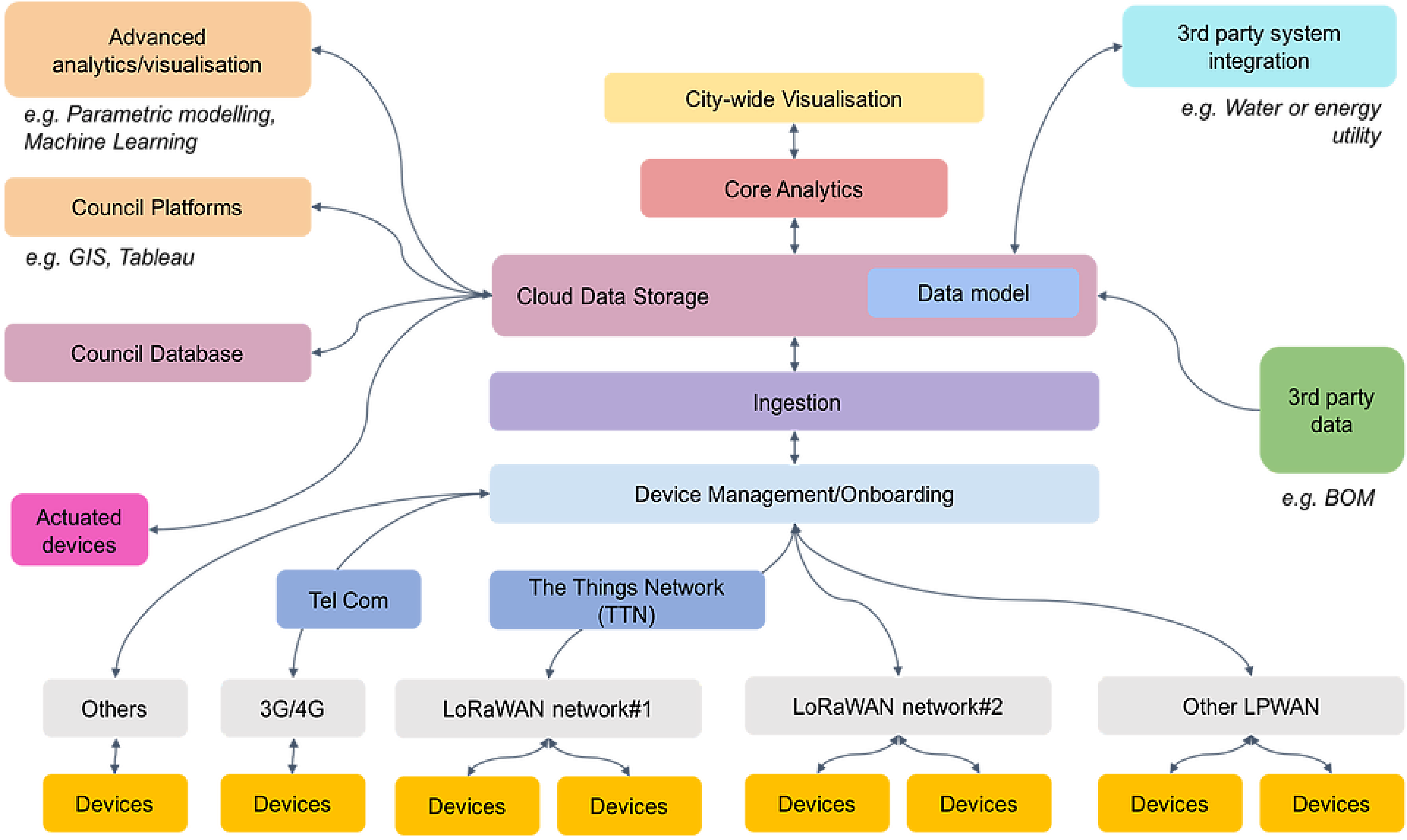

**Figure 4.** The overall architecture of the ROSE architecture (with the kind permission from the IoT Australia).

An assessment, here in ROSE, was performed via the IoT-PA prototype version of our framework. The assessor could select the most important criteria she/he wanted to check against ROSE (Figure 5a). The provided leading questions related to all the layers in the IoT-PA would allow the assessor to qualitatively express the level of satisfaction of the criteria (Figure 5b). For each criterion, the assessor could examine if ROSE has defined supportive mechanisms or techniques in addressing that criterion. For example, for the criteria generating meta-data (Section 4), the IoT-PA asks the assessor if ROSE defines mechanisms for annotating the data being collected from smart objects, users, and applications. Satisfying this criterion was important since it could result in the improved data classification and query processing. The IoT-PA defines seven Likert scales (1–7) for the rating of the criteria satisfaction where 1 represents not at all addressed, 2 for partially addressed, 3 for somewhat addressed, 4 for neutral, 5 for mostly addressed, 6 for considerably addressed, and 7 for completely addressed. For each criterion, if there is more than one leading question, the rated answers for the questions are aggregated. The assessor could use a voting mechanism including all the ROSE developers' opinions to accurately estimate the extent to which a criterion is satisfied in the ROSE architecture. Figures 5b and 5c, respectively, show the ratings of the criteria satisfaction provided by the assessor and the evaluation results for ROSE. The evaluation exercise was iterative and evolutionary in the sense that the current evaluation results were fed into the next design iteration to refine the architecture. Thus, the next round of the evaluation exercise could be based on the refined architecture from the prior design iteration. The evaluation exercise results could be stored in the IoT-PA database and updated during the design episode.

The assessment scenario had some lessons learned and opportunities to revise our framework toward its next iteration of the DSR. The senior research consultant of the IoT Australia confirmed our frameworks' adherence to the meta-criteria MC4, i.e., soundness and MC9, i.e., comprehensiveness as he believed that the proposed framework criteria are important and relatively sufficient for the incorporation into the ROSE architecture design. Regarding the meta-criterion MC5, i.e., generality, an area of concern that was raised by the technical lead of ROSE was that he believed the criteria should be contextualized for a specific domain along with the evidential examples. That is, the criteria specialization should be derived from the project context and domain-specific functions of a platform. For example, there should be criteria related to the device management examining how aspects such as metadata generation, data ingestion, or device integration are addressed in the architecture. Addressing this comment, i.e., a use-case centric evaluation, yet addressable by adding new leading questions to our framework, indeed, was against the meta-criteria MC1, MC2, and MC5 as defined in the design phase. These were not supported by the framework. We believed that this deficiency in our framework was rooted in the changing requirements of IoT Australia, i.e., a change from having a generic evaluation framework to a domain-specific one. This is further discussed in Section 6.2. We did not receive noticeable comments from the IoT Australia regarding other meta-criterion MC3 (minimum overlapping), MC6 (consistency), MC7 (balance), and MC8 (measurability).

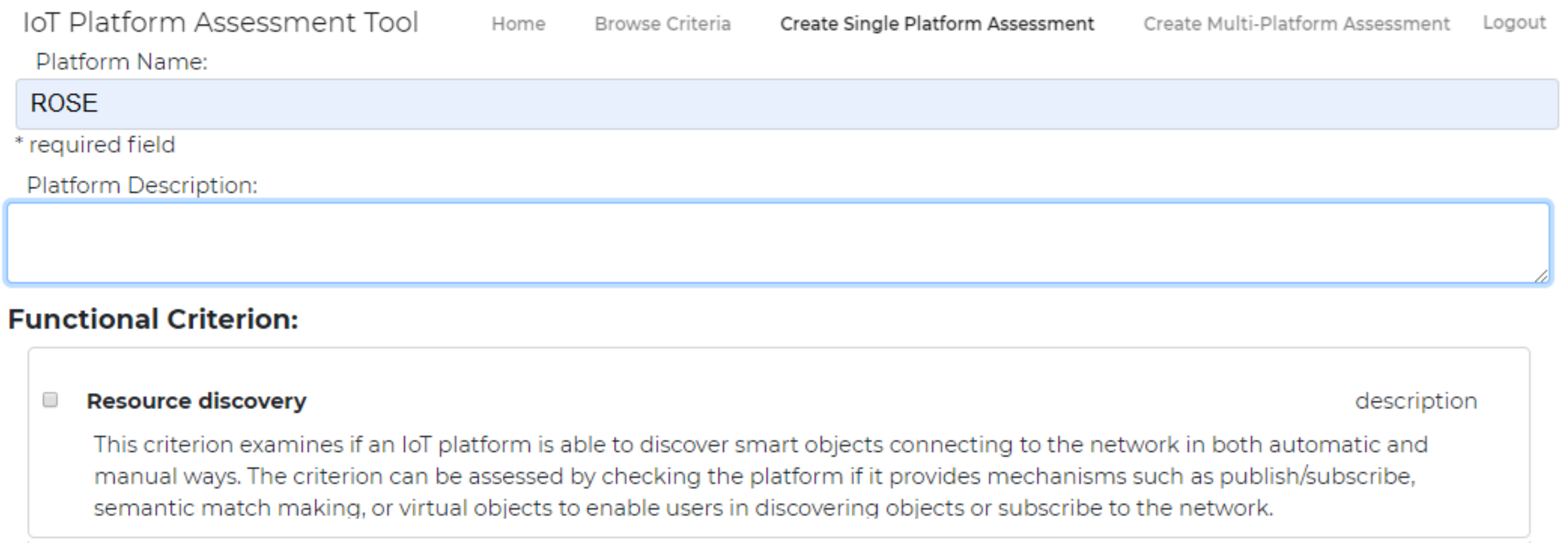

**(a)** sample description of the framework criteria

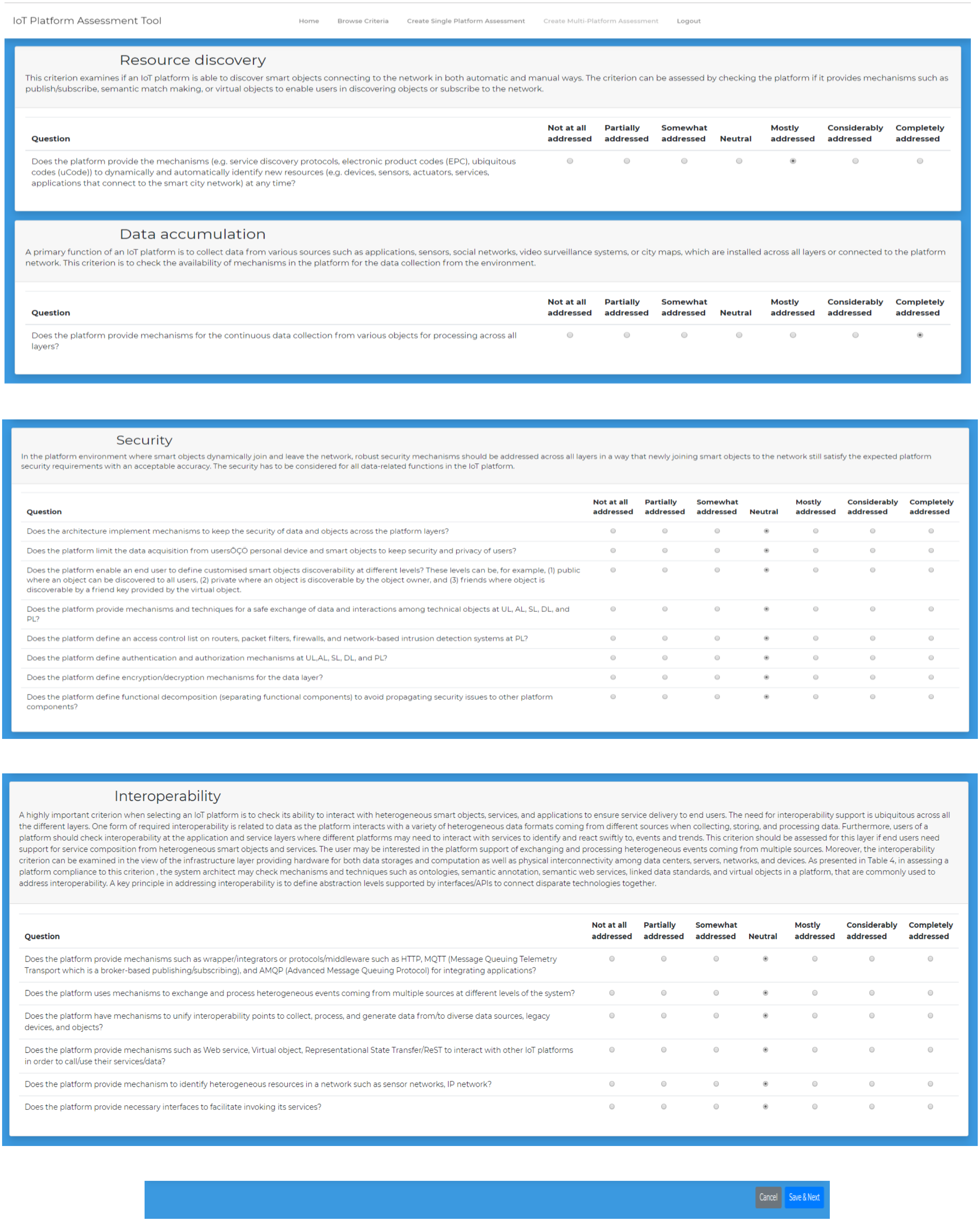

**(b)** sample rates per criterion

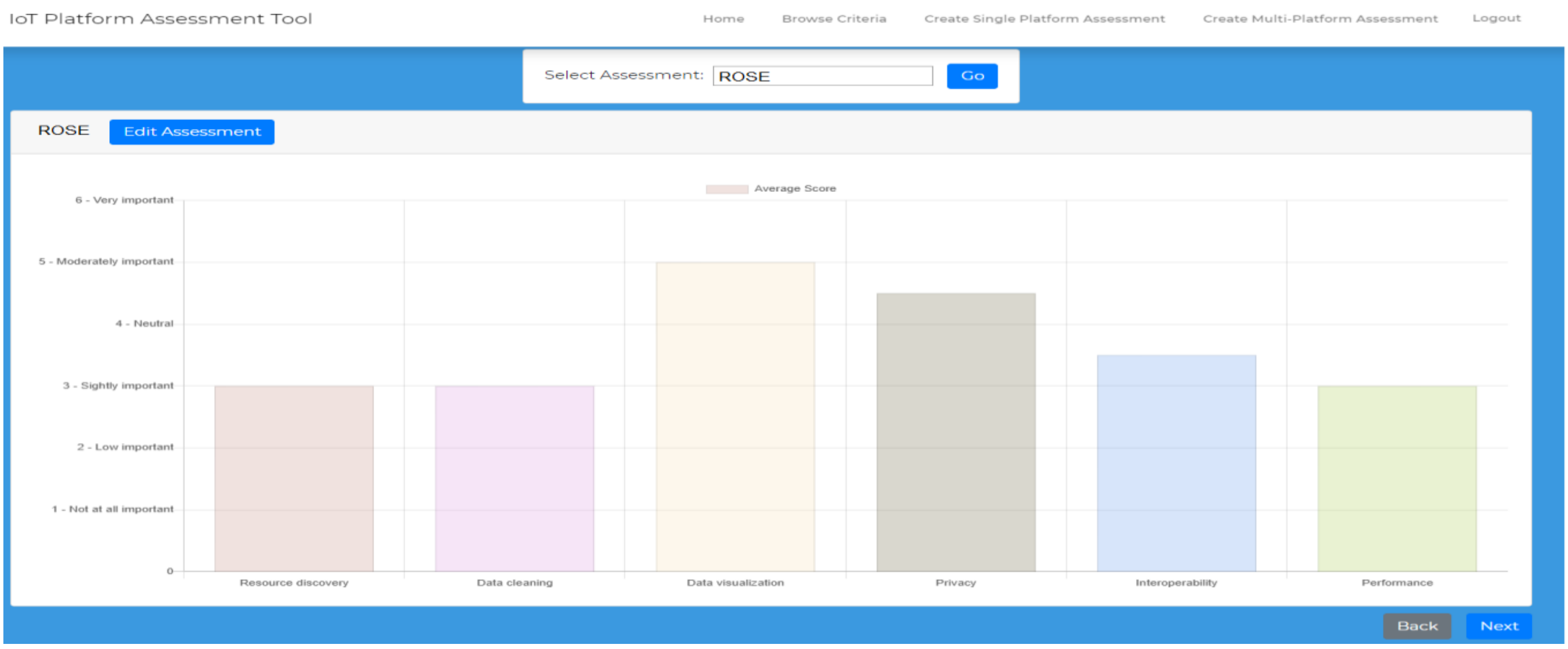

(**c**)sample descriptive stats showing the satisfaction level of criteria

**Figure 5.** Assessment of the ROSE architecture via the IoT-PA (the rates are exemplar due to confidentiality).

### *5.2. Case Study 2: Multi-Platform Analysis*

Deciding which IoT platform meets the requirements entails devising approaches that incorporate multiple criteria and their different levels of importance. Research on the evaluation and selection of products falls under the decision science research stream, in particular, multi-criteria decision-making (MCDM) problem [21]. A critical analysis of the criteria and their relationships to each other is necessary as it impacts the overall ranking and selection of platforms. There are different techniques to solve MCDM problems. Amongst them, AHP is one of the common techniques for solving MCDM problems. The AHP procedure [22] was first introduced by Thomas L. Saaty and later has been widely adopted by many researchers and practitioners in several areas including the selection of business maturity model [23], online advertisement websites [24], product selection [25], software models selection [26], and many others. The AHP, typically, is used in technology adoption decision-making scenarios by characterizing multiple criteria and a predetermined set of decision alternatives. User judgments are the input source to conduct the AHP. These judgments are described as pairwise comparisons of decision criteria. Pairwise comparisons are based on the relative importance of one criterion over another in addressing a specific goal. This relativity makes AHP less sensitive to judgmental errors. Each pairwise of comparison results in a numerical value index (i,j) representing the estimate of the ratio between the weights of the two criteria indexed by $i$ and $j$. Analyzing a variety of IoT platforms as each offers a different set of features against a set of criteria with different priorities is a challenging exercise that can be addressed using the AHP.

#### 5.2.1. AHP Procedure

Step 1. Structuring the decision hierarchy. It might be the case that all the criteria are not concerned for the evaluation purpose, i.e., a shortlisting of criteria depends on the factors such as stakeholders' priorities or an application domain. For example, in a domain such as aviation systems, addressing the criteria such as recoverability by a platform may have a higher priority over security and thus should be taken into account. Conversely, the privacy criterion may take precedence over the platform recoverability in the case of public service for citizens.

Step 2. Comparing criteria and weighing their relative importance. Due to the fact that assessors may have different preferences regarding the priority of the criteria, the relative weights/importance of each criterion $c_i$ ($1 \le i \le k$) should be determined. For $k$ criteria, the assessor needs to perform *k*(k–1)/2* pairwise comparisons. Although the assessor may use his/her own weights for the pairwise comparison, our framework borrows the original weights provided by the AHP as shown in Table 5. By comparing each pair of the criteria in a weighing scale in Table 5, the assessor collectively provides her preference on the relative importance of the criteria. For instance, if $c_i$ is strongly preferred over $c_j$, ($1 \le i, j \le k$), then the entry of comparison matrix ccm(*i,j*) = 5 and symmetrically ccm(*j*,i) = 1/5 as shown below.

$$\text{Criteria comparison matrix (ccm)} = \begin{bmatrix} 1 & \dots & c_{i,j} = 5 \\ \dots & 1 & \dots. \\ c_{j,i} = 1/5 & \dots & 1 \end{bmatrix}$$

Based on ccm, the priority vector of the criteria is computed. A common technique to estimate the priority vector is the geometric mean proposed by Crawford and Williams [27] where each element of the priority vector is assigned with the geometric mean of the elements on the respective row divided by a normalization term. The summation of the elements of the priority vector is equal to 1. This is shown in Equation (1):

$$\text{Weight } (c_i) = (\prod_{j=1}^{k} ccm_{i,k})^{1/k} / \sum_{i=1}^{k} (\prod_{j=1}^{k} ccm_{i,k})^{1/k} \quad (1)$$

Thus, the criteria priority vector (cpv):

$$\text{cpv} = [\text{Weight } (c_1), \dots, \text{Weight } (c_k)\ ]$$

The priority vector *cpv* shows the relative weights among the criteria are being comparing.

**Table 5.** Pairwise comparison of item X and item Y.

| Qualitative Judgment | Numerical Rating |
|---|---|
| X is equally preferred to Y | 1 |
| X is equally to moderately preferred over Y | 2 |
| X is moderately preferred over Y | 3 |
| X is moderately to strongly preferred over Y | 4 |
| X is strongly preferred over Y | 5 |
| X is strongly to very strongly preferred over Y | 6 |
| X is very strongly preferred over Y | 7 |
| X is very strongly to extremely preferred over Y | 8 |
| X is extremely preferred over Y | 9 |

Step 3. Weighing IoT platforms in addressing the criteria. All platforms are compared against each other based on each criterion. Again, using the qualitative values presented in Table 5, the assessor compares all *m* platforms regarding each criterion $c_i$ in a pairwise manner. This comparison indicates how platform $p_a$ supports the criterion $c_i$ compared to platform $p_b$ ($1 \le a, b \le m$). For instance, if $p_a$ is strongly preferred over $p_b$, in the comparison matrix there should be an entry $p_a$, $p_b$ = 5 and symmetrically $p_b$, $p_a$ = 1/5 as shown below.

$$\text{Platform comparison matrix (pcm)} = \begin{bmatrix} 1 & \dots & p_{i,j} = 5 \\ \dots & 1 & \dots. \\ p_{j,i} = 1/5 & \dots & 1 \end{bmatrix}$$

The comparison results in new comparison matrices for each criterion. Similar to the previous step, the priority vectors for each matrix table is created using Equation (1). Obviously, in Equation (1), the index *k* which indicates the number of criteria should be replaced with the *m*, i.e., the number of platforms as shown in Equation (2):

$$\text{Weight } (p_i) = (\prod_{j=1}^{m} pcm_{i,k})^{1/k} / \sum_{i=1}^{k} (\prod_{j=1}^{k} pcm_{i,k})^{1/k} \quad (2)$$

Step 4. Computing the overall value of an IoT platform assessment. The overall composite weight of each alternative IoT platform is computed based on the priority vectors produced in steps 2 and 3. This is computed using Equation (2) as follows:

$$Composite\ \ weight\ \ p_i = \sum_{i=1}^{k} \text{Weight}\ \ (p_i) * \ \ \text{Weight}\ \ (c_i) \tag{3}$$

The highest value score signifies the most suitable platforms matching user's preferences on the criteria.

### 5.2.2. Example Scenario

The applicability of the framework in comparing and ranking given example IoT platforms is discussed in this section. The example IoT platforms were the AWS IoT Platform, IBM Watson IoT Platform, and Microsoft Azure IoT Platform, respectively, denoted by $p_1,\ p_2, and\ \ p_3$. They are being widely used as a benchmark for both academia and practitioners in the IoT field. The judged weights in the comparison matrix have been based on our knowledge and obtained data through reviewing the existing documents of the platform conducted in the design phase of the framework design (Section 3.1). The following presents the framework's steps for the ranking of these platforms.

The IoT-PA prototype was a useful tool to automate the calculations of the AHP procedure. The assessor was interested in two functional-related criteria query processing ($c_1$) and data visualization ($c_2$) and three non-functional criteria security ($c_3$), reusability ($c_4$), and extensibility ($c_5$). In the second step of the AHP procedure, the pairwise comparison matrix of the criteria was performed, and their relative weights were computed as shown in Figure 6.

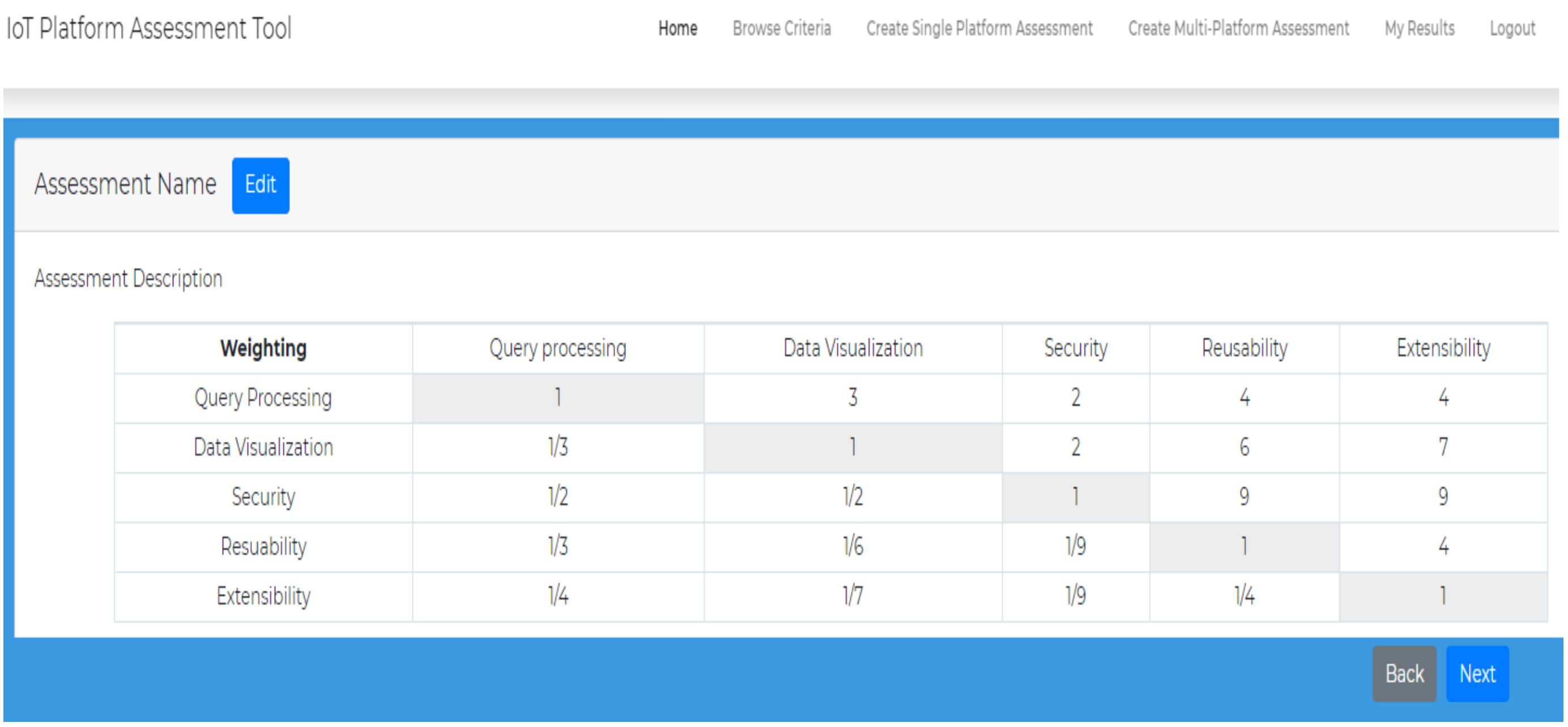

| Weighting | Query processing | Data Visualization | Security | Reusability | Extensibility |
|---|---|---|---|---|---|
| Query Processing | 1 | 3 | 2 | 4 | 4 |
| Data Visualization | 1/3 | 1 | 2 | 6 | 7 |
| Security | 1/2 | 1/2 | 1 | 9 | 9 |
| Resuability | 1/3 | 1/6 | 1/9 | 1 | 4 |
| Extensibility | 1/4 | 1/7 | 1/9 | 1/4 | 1 |

**Figure 6.** Pairwise comparison of the selected criteria.

Next, the priority vector was computed for the criterion using Equation (1). For example, the relative weight for the interoperability ($c_1$) was:

$$\text{Weight}\ \ (c_1) = (1*3*2*4*4)^{1/5} / \left((1*3*2*4*4)^{1/5} + (1/3*1*2*6*7)^{1/5} + (1/2*1/2*1*9*9)^{1/5} + (1/4*1/7*1/9*1/4*1)^{1/5}\right)\ \ = \ 0.35$$

Similarly, the priority vector can be computed for portability ($c_2$):

$$\text{Weight}\ \ (c_2) = (1/3*1*2*6*7)^{1/5} / \left((1*3*2*4*4)^{1/5} + (1/3*1*2*6*7)^{1/5} + (1/2*1/2*1*9*9)^{1/5} + (1/4*1/7*1/9*1/4*1)^{1/5}\right)\ \ = \ 0.27$$

Following the similar computations for other criteria security ($c_3$), reusability ($c_4$), and extensibility ($c_5$), the priority vector of the criteria is:

$$\text{cvp} = [0.35, 0.27, 0.26, 0.06, 0.05]$$

The relative weight of the platforms for each criterion was computed through step 3. For each criterion, a comparison matrix was created. This was followed by computing the priority vectors for each criterion using Equation (2) as shown in Figure 7.

IoT Platform Assessment Tool — Home | Browse Criteria | Create Single Platform Assessment | Create Multi-Platform Assessment | My Results | Logout

| | | AWS | IBM | Azure | Priorities |
|---|---|---|---|---|---|
| Query Processing | AWS | 1 | 1/4 | 3 | 0.24 |
| | IBM | 4 | 1 | 4 | 0.65 |
| | Azure | 1/3 | 1/4 | 1 | 0.11 |
| Data Visualization | AWS | 1 | 2 | 6 | 0.58 |
| | IBM | 1/2 | 1 | 5 | 0.34 |
| | Azure | 1/6 | 1/5 | 1 | 0.08 |
| Security | AWS | 1 | 4 | 6 | 0.69 |
| | IBM | 1/4 | 1 | 3 | 0.22 |
| | Azure | 1/6 | 1/3 | 1 | 0.09 |
| Resuability | AWS | 1 | 2 | 7 | 0.62 |
| | IBM | 1/2 | 1 | 3 | 0.29 |
| | Azure | 1/7 | 1/3 | 1 | 0.09 |
| Extensibility | AWS | 1 | 2 | 1/4 | 0.47 |
| | IBM | 1/2 | 1 | 4 | 0.38 |
| | Azure | 4 | 1/4 | 1 | 0.15 |

**Figure 7.** Priority vectors for the IoT platforms.

For example, the weight of the platform AWS with respect to the criterion interoperability ($c_1$) was computed using Equation (2):

$$\text{Weight } (p_1) = (1 * 1/4 * 3)^{1/3} / \left((1 * 1/4 * 3)^{1/3} + (4 * 1 * 4)^{1/3} + (1/3 * 1/4 * 1)^{1/3}\right) = 0.24$$

Finally, we computed the overall composite weight for each alternative platform based on the weight of the criteria using Equation (3). The overall weight was the linear combination of priority vectors of criteria and platforms as follows.

Overall platform ranking (opr) =

$$\begin{pmatrix} \text{AWS} & 0.58 & 0.58 & 0.69 & 0.62 & 0.47 \\ \text{IBM} & 0.30 & 0.34 & 0.22 & 0.29 & 0.38 \\ \text{Microsoft} & 0.12 & 0.08 & 0.09 & 0.09 & 0.15 \end{pmatrix} \times \begin{pmatrix} \text{query processing} & 0.35 \\ \text{data visualisation} & 0.27 \\ \text{security} & 0.26 \\ \text{reusability} & 0.06 \\ \text{extensibility} & 0.05 \end{pmatrix} = \begin{pmatrix} \text{AWS} & 0.598 \\ \text{IBM} & 0.272 \\ \text{Microsoft} & 0.1067 \end{pmatrix}$$

Therefore, the ranking of the IoT platforms can be decided based on the matrix *opr*. That is, AWS is relatively the best choice following by IBM as the second choice and Microsoft as the third choice. In other words, it can be interpreted that the AWS platform is 0.59 times more preferable than IBM. The overall platform ranking vector provides an aggregative view of the criteria. It is possible to visualize the differences of IoT platforms in addressing the criteria. This can be shown using the Kivait diagram. As shown in Figure 8, it seems that IBM provides the best support for query processing ($c_1$). This is followed by the AWS and Microsoft IoT platforms. In terms of criterion security ($c_3$), AWS provides the best support compared to other platforms. This implies that if smart city services require security guaranteed, then AWS is the most suitable choice in terms of security. On the other hand, in terms of data visualization ($c_2$), AWS provides better support for data representation.

The subjectivity might be an issue since the abovementioned framework evaluations have been performed by the authors of this paper. Depending on the objectives and the context of an evaluation, alternative qualitative and quantitative techniques may be suited. Qualitative techniques such as Delphi can be adopted where a panel of experts evaluate IoT platforms of AWS, IBM Watson, and Microsoft Azure. Any significant differences in the ratings are then resolved via post-evaluation discussions to reach a consensus. On the other hand, empirical techniques such as experiments or stats about IoT platforms can be used for evaluation.

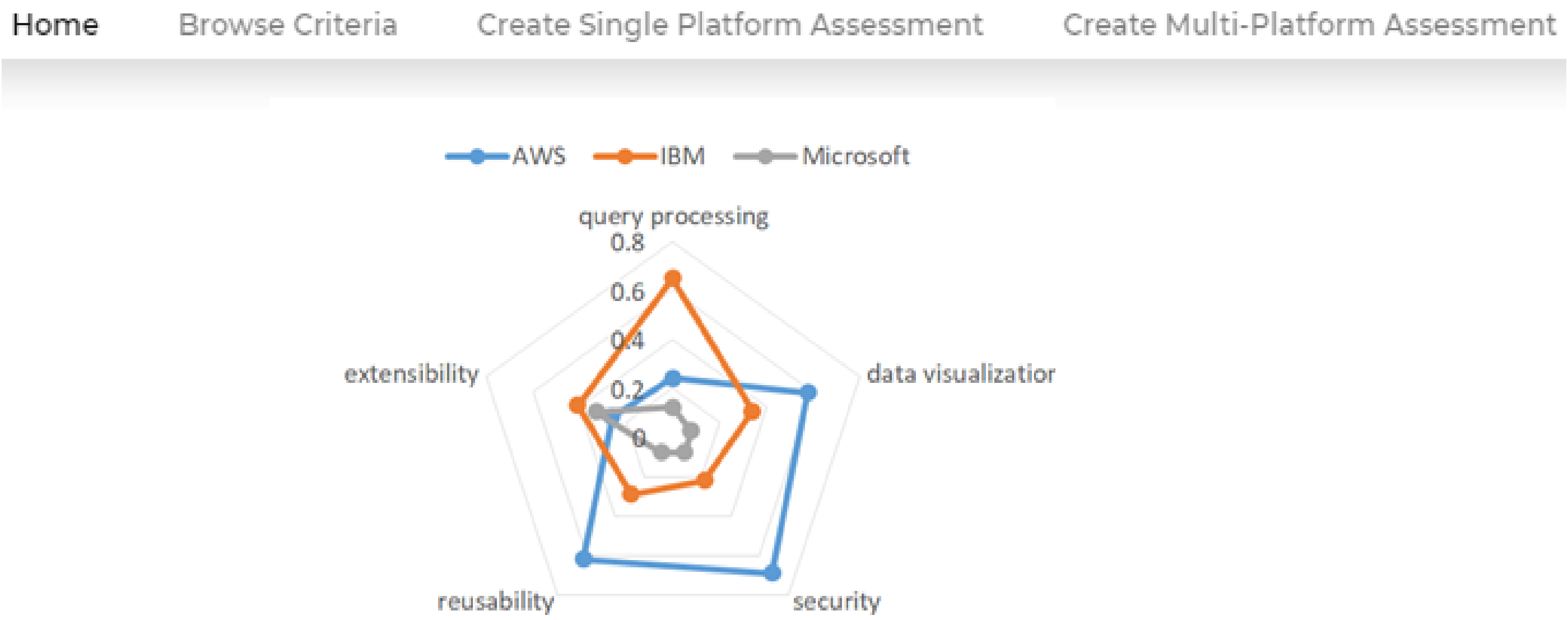

Figure 8. Comparison of IoT platforms for the criteria query processing ($c_1$) and data visualization ($c_2$) and three non-functional criteria security ($c_3$), reusability ($c_4$), and extensibility ($c_5$).

## 6. Discussion

### *6.1. Research Implications*

This work contributes to the research and practice by filling the gap of IoT platform assessment and unavailability of a DSR approach in the IoT Australia context. The following discusses the implications of our work to research, practice, and the DSR.

Implications to research and practice. Our framework has been grounded on the well-established notion of software architecture from the software engineering discipline. The software architecture is an abstraction of a target system, and it manifests further phases of system development and maintenance. Such an architecture-centric view to IoT platform architecture design was adopted to identify and compile a collection of relevant and most important evaluation criteria. Researchers can be reasonably confident that our framework has been built up upon an established way, i.e., software architecture. The criterion set can be incorporated into the early-stage analysis of an IoT platform architecture in a white-box fashion or multiple-platform comparison via MCDM techniques such as the AHP in a black-box way. This, in turn, allows researchers to explore in-depth the quality of the architecture design for their IoT-based solutions per criterion. Being agnostic is a key feature of our framework that makes it independent of underlying IoT technologies and domain-specific IoT platforms.

Since the traditional AHP cannot be effectively applied for analyzing IoT platforms as it leaves the criteria definition to assessors, a contribution of our work is to extend the AHP to the IoT domain. Using the AHP procedure is not unique in our research; however, specializing the AHP via well-established criteria and leading questions makes it more applicable to the IoT domain.

Apart from the IoT domain, due to the importance of adherence to the proposed criteria and their relevance to other corresponding technologies, i.e., IoT enabling technologies such as the cloud computing and data analytics platforms, the framework can be extended to different domains of technology selection when assessors are interested in evaluating possible alternatives.

In terms of practical implications, an organization may have its own in-house framework for the suitability assessment of embarking on new technologies in organizational business processes, but the framework may have deficiencies in supporting IoT adoption scenarios. Managers can incorporate our proposed framework into their existing assessment framework portfolios to conduct decision-making scenarios with a focus on the criteria attuned with the IoT domain.

Implications to DSR. Firstly, our DSR project was based on a formal contract between the IoT Australia and the UTX researchers. This committed both project parties to have formal/informal meetings, participate in workshops, information exchange, i.e., the ROSE architecture, documentation, and analyze requirements toward developing the target framework. We perceived that the active stakeholder participation during the DSR endeavor is critical for effectively developing IT artefact for a stated problem. The active engagement of stakeholders, as it is also acknowledged by Asif et al. [28] in adopting the action DSR for an information system architecture configuration project, was difficult to achieve in our project. This due to different reasons. For example, both parties had different project time frames, priorities, and preferences to appoint meetings/interviews because of other daily responsibilities. This implies having countermeasures to tackle issues in the active stakeholder engagement in conducting DSR projects.

Secondly, we experienced the changes in IoT Australia requirements during the project that, unavoidably, were propagating our DSR process of framework artefact design. For example, at the early stages of the project, the IoT Australia called the UTX team to design a framework to assess IoT-based smart city architectures. Nevertheless, this request later turned to design a framework to assess IoT platforms as an underlying enabler of smart city services. Additionally, the IoT Australia initially requested the UTX researchers to provide a generic evaluation framework that facilitates high-level IoT platform architecture, i.e., ROSE. This request turned to a need for a use-case-centric and domain-specific evaluation framework. This was in contradiction with the adherence to the meta-criterion MC5, i.e., generality, which was in the design phase of the DSR project. These examples imply the fact that researchers of the DSR projects need to anticipate change management mechanisms as they may have limited resources and fixed timeframe, i.e., deadlines for submitting research outcomes to international venues. This is an important practical concern that has not been well highlighted in DSR literature.

Thirdly, a DSR project might be aimed at a co-creation IT artefact development with mutual benefits to all involved parties who may join the project with different overall objectives, conflicting priorities, and working horizons. Perhaps, two main artefacts from these projects have been an executable system/framework/tool for IoT Australia and academic publications for UTX researchers. These two types of artefacts as the project timeframes and objectives of the IoT Australia and UTS were different need to proper planning. Making a balance between the priorities of both parties and mitigating the pitfall of turning a DSR project into a full industry-oriented consultation or pure research-oriented endeavor are the important managerial implications in conducting DSR projects.

### *6.2. Research Limitations*

We have tended to capture commonly occurred criteria in the existing IoT literature in our framework. Unavoidably, there might be some less commonly cited criteria, however important, but our framework does not cover them. We do not claim the inclusivity of the framework criteria for different platform evaluation scenarios. This confines the generalizability of the proposed criteria in the view of the meta-criteria if they are to be used in domain-specific assessment scenarios.

The development of the initial set of 16 non-functional-related criteria has been our first step toward constructing IoT platform architecture evaluation framework. The base framework needs to cover further criteria perceived as highly important, generally or specifically, to a particular paradigm. One of which, for instance, is that the framework may need the extension to cater to security and privacy-related criteria specific to a domain. In addition, the compliance with local legislation, e.g., GDPR (General Data Protection Regulation) in Europe, requires further framework enhancement. The criteria that are related to licensing, i.e., mechanisms indicating the availability of hardware and software components for on-premise or as a service, may also need to be considered. These will be the subjects of our future work.

Moreover, IoT platforms rely on enabling technologies such as cloud computing and data analytics. In other words, an IoT platform evaluation exercise may be extended to the assessment of third-party services. For example, if the platform data is stored on a cloud data storage, then the security mechanism provided to select the cloud storage is assessed. Following the DSR approach, the above limitations can be considered as new increments to the framework over the next iteration of the DSR.

### *6.3. Related Work*

There are several relevant evaluation frameworks supporting the feature analysis in the software architecture design arena such as SAAM (software architecture analysis method) [29], ATAM (architecture trade-off analysis method) [7], CBAM (cost–benefit analysis method) [7], ALMA (architecture level modifiability analysis) [30], and FAAM (family–architecture analysis method) [31]. For instance, ATAM is a promulgated scenario-based software architecture evaluation framework at the presence of interactions and interdependencies among quality attributes. It highlights trade-off mechanisms and opportunities between different qualities. We believe the leading questions for the criterion set in our framework can be incorporated as a complementary step into the frameworks noted above to enhance their capability in evaluation support of IoT-specific characteristics.

Although the analytical evaluation and ranking of various IoT platforms is almost an entirely new topic in the IoT research area, the main idea has been extensively used in other closely related fields such as service computing and cloud computing. There are synergies between Internet-based computing paradigms such as cloud computing and IoT as they share similar characteristics and provide reciprocal backbone services[32]. For instance, cloud computing provides resources for the storage and distributed processing of the acquired sensor data in an IoT platform. In the cloud-computing literature, for instance, the works by Garg et al. [33] and Khajeh-Hosseini et al. [34], jointly, concern with the problem of cloud service selection where they have developed a set of metrics to find suitable infrastructure as service (IaaS) platforms. However, the coverage of the criteria is insufficient to be used in the context of IoT platform assessment. None of the criteria related to the core IoT platform functions such as resource discovery, data accumulation, data cleaning, data storing, data analysis, and query processing are supported by their works. The second concern related to this class of frameworks is the insensitivity of their criteria definitions. In the IoT context, a criterion may have different interpretations across the architecture layers; however, this distinction has not been a feature of the proposed criterion set in [33,34]. For instance, the interoperability criterion in our framework is subject to the architecture layers of IoT platforms such as data, service, application, and physical. None of these layers are defined in [33,34]. Furthermore, the abovementioned works do not cover the criterion such as mobility, reusability, extensibility, and maintainability that we have included in our framework.

Pertinent to the IoT literature, there is a dearth of studies aiding users in assessing and comparing existing IoT platforms. A large group of studies, such as [12,35–37], to name a few, are literature surveys presenting a comparative analysis of IoT platforms, but, providing guidelines for an assessor to evaluate platforms, compare, and rank has not been the focus. In contrast, our work is complementary to these literature surveys since it adds more rigor to measuring criteria by providing leading questions. Choi et al. [38] suggested an approach to assess the priority of criteria such as stability, productivity, and diversity in business processes utilizing IoT initiatives. The work asks the perceptions of 51 IoT researchers on the analytical criteria. The main difference between our framework and [38] is that we narrowed our focus to an IoT platform architecture as the unit of analysis and develop criteria applicable to evaluate and rank different given platform alternatives. The work presented in [39] assesses security criterion issues related to IoT adoption, but it confines its view to security criterion to compare IoT platforms at the network layer such as open port, traffic, and CPU. Unlike [39], we expand the evaluation across the layers of a typical IoT platform stack. The presented framework in this article extends our earlier work in [40] where we conducted an initial evaluation of nine existing IoT architecture examples.

In the work by Kondratenko et al. [41], the authors are concerned with increasing the accuracy of decision making; they suggest using two different methods of multi-criteria decision making, for instance, the linear convolution and multiplicative convolution, to get to know which platform may suit the criteria. The work by Kondratenko et al. [41] including other studies[42,43] provides ad hoc quantitative techniques, albeit costly and complex, for developing quantifiable criteria to increase the reliability of decision outcomes in an IoT platform selection scenario. It remains that [41–43] have not provided criteria and guidelines examining the suitability of an IoT platform for a given problem domain. Our work is in contrast to these quantitative techniques relying on the subjective definition of criteria that have no IoT-specific meaning and can therefore be difficult to use in a real evaluation exercise. Our framework can be used as a complementary step to consolidate the criterion list and be used as an input for the above studies. The important IoT platform provider and consumer's question of what essential criteria should be taken into account when designing or selecting an IoT platform has remained unanswered and is indeed our driving motivation in this paper.

## 7. Conclusions

The growth in the number of IoT platform variants has raised the need for evaluation and comparison efforts, mostly in order to assist in appraising an existing platform architecture, selecting, or designing a new one aimed at satisfying specific quality criteria. The availability of such a framework is non-extant in the literature. In this context, this paper demonstrated the application of the DSR along with kernel theories, extant artefacts, and industry advice to design and validate a framework for IoT platform architecture evaluation in a one year industry project. The framework aimed at enabling IoT Australia to assess the quality of IoT-based solution architecture in the municipal city council at NSW. The framework has been underpinned on a commonly grounded set of architecture-centric functional and non-functional criteria by which an ideal IoT platform is expected to adhere to. The rationale behind designing the framework has been to provide a basis for a fine-grained and in-depth evaluation and comparison of IoT platforms. The novelty of our framework is hence not merely to provide an established and more comprehensive criterion set but also incorporating these criteria into the AHP for providing a quality-aware comparison and ranking of platforms. The framework is the first attempt to enable users to systematically assess different layers of platforms and can be viewed as a quality guideline for both researchers and practitioners who are developing and maintaining IoT platforms.

**Author Contributions:** "Conceptualization, Mahdi Fahmideh; methodology, Mahdi Fahmideh; software, Mahdi Fahmideh; validation, Mahdi Fahmideh; formal analysis, Mahdi Fahmideh.; investigation, Mahdi Fahmideh; resources, Mahdi Fahmideh; data curation, Mahdi Fahmideh; writing—original draft preparation, Mahdi Fahmideh; writing—review and editing, Mahdi Fahmideh, Jun Yan, Jun Shen, Davoud Mougouei, Yanlong Zhai, and Aakash Ahmad; visualization, Mahdi Fahmideh, Aakash Ahmad; supervision, Mahdi Fahmideh; project administration, Mahdi Fahmideh. All authors have read and agreed to the published version of the manuscript."

**Funding:** This project was a part of the urban livability and smart cities project funded by the Australian Federal Government Smart Cities and Suburbs program, Department of Industry, Innovation and Science.

**Institutional Review Board Statement:** Not applicable

**Informed Consent Statement:** Not applicable

**Data Availability Statement:** Not applicable as the study does not report any data.

**Acknowledgments.** The authors acknowledge the support and guidance received from IoT Alliance Australia at University of Technology Sydney, Australia. Finally, the authors would also like to thank Dr. Didar Zowghi, the professor of software engineering in the Faculty of Engineering and IT, University of Technology Sydney, for her constructive comments and suggestions in the first draft of this article.

**Conflicts of Interest:** The authors declare no conflict of interest.

## Appendix A

**Literature Source Used to Develop the Framework Evaluation Criteria**

| ID | Authors and Title | Acronym | Source | Year |
|---|---|---|---|---|
| [S1] | Eduardo Santana, Zambom Felipe, et al., "Software platforms for smart cities: Concepts, requirements, challenges, and a unified reference architecture" | RASCP | ACM Computing Surveys | 2017 |
| [S2] | Riccardo Petrolo, Valeria Loscri, et al., "Towards a Smart City based on Cloud of Things, a survey on the smart city vision and paradigms" | VITAL | IEEE | 2014 |
| [S3] | Bin Cheng, Salvatore Longo, et al., "Building a Big Data Platform for Smart Cities: Experience and Lessons from Santander" | CiDAP | IEEE | 2015 |
| [S4] | Zaheer Khan, Ashiq Anjum, et al., "Cloud Based Big Data Analytics for Smart Future Cities" | - | IEEE/ACM | 2013 |
| [S5] | Jayavardhana Gubbi, Rajkumar Buyya, et al., "Internet of Things (IoT): A Vision, Architectural Elements, and Future Directions" | - | Elsevier | 2013 |
| [S6] | Cisco, "The Internet of Things Reference Model" | Cisco | Cisco | 2014 |
| [S7] | Federico Ciccozzi, Crnkovic Ivica, "Model-driven engineering for mission-critical IoT systems" | MC-IoT | IEEE | 2017 |
| [S8] | Sebastian Lange, Andreas Nettsträter, et al., "Introduction to the architectural reference model for the Internet of Things" | IoT-ARM | IoT-A | 2013 |
| [S9] | John Soldatos, Nikos Kefalakis, "OpenIoT: Open source Internet-of-Things in the cloud" | OpenIoT | Springer | 2015 |
| [S10] | Jasmin Guth, Uwe Breitenbucher, et al., "Comparison of IoT Platform Architectures: A Field Study based on a Reference Architecture" | - | IEEE | 2016 |
| [S11] | Ivan Ganchev, Zhanlin Ji, et al., "A Generic IoT Architecture for Smart Cities" | - | IEEE | 2014 |
| [S12] | "FIWARE (also called Open & Agile Smart Cities (OASC))" | FIWARE/OASC | FIWARE Community | 2014 |
| [S13] | Ignasi Vilajosana, Jordi Llosa, et al., "Bootstrapping smart cities through a self-sustainable model based on big data flows" | - | IEEE | 2013 |
| [S14] | Kohei Takahashi, Shintaro Yamamoto, et al., "Design and implementation of service API for large-scale house log in smart city cloud" | Scallop4SC | IEEE | 2012 |
| [S15] | Zaheer Khan, Ashiq Anjum, et al., "Towards cloud based big data analytics for smart future cities" | - | Springer Open Journal | 2015 |
| [S16] | Catherine E. A. Mulligan, Magnus Olsson, "Architectural Implications of Smart City Business Models: An Evolutionary Perspective" | - | IEEE | 2013 |
| [S17] | George Kakarontzas, Leonidas Anthopoulos, et al. "A Conceptual Enterprise Architecture Framework for Smart Cities, A Survey Based Approach" | EADIC | IEEE | 2014 |
| [S18] | David Díaz Pardo de Vera, Álvaro Sigüenza Izquierdo, et al. "A Ubiquitous sensor network platform for integrating smart devices into the semantic sensor web" | Telco USN-Platform | Sensors | 2014 |
| [S19] | Sotiris Zygiaris, "Smart City Reference Model: Assisting Planners to Conceptualize the Building of Smart City Innovation Ecosystems" | SCRM | Springer | 2012 |
| [S20] | Dennis Pfisterer, Kay Romer, "SPITFIRE: Towards a Semantic Web of Things" | SPITFIRE | IEEE | 2011 |
| [S21] | Nam K Giang, Rodger Lea, et al., "On Building Smart City IoT Applications: a Coordination-based Perspective" | - | ACM | 2016 |
| [S22] | Rong Wenge, Xiong Zhang, et al., "Smart City Architecture: A Technology Guide for Implementation and Design Challenges" | - | IEEE | 2014 |
| [S23] | Paul Fremantle, "A reference architecture for the internet of things" | WSO2 | WSO2 | 2015 |
| [S24] | Andrea Zanella, Senior Member, "Internet of Things for Smart Cities" | Padova | IEEE | 2014 |
| [S25] | Wolfgang Apolinarski, Umer Iqbal, et al., "The GAMBAS Middleware and SDK for Smart City Applications" | GAMBAS | IEEE | 2014 |
| [S26] | Nader Mohamed, Jameela Al-Jardoodi, "SmartCityWare: A Service-Oriented Middleware for Cloud and Fog Enabled Smart City Services" | SmartCityWare | IEEE | 2017 |
| [S27] | Jiong Jin, Jayavardhana Gubbi, "An information framework for creating a smart city through internet of things" | Noise mapping | IEEE | 2013 |
| [S28] | Panagiotis Vlacheas, Vera Stavroulaki, et al., "Enabling Smart Cities through a Cognitive Management Framework for the Internet of Things" | - | IEEE | 2013 |
| [S29] | Aditya Gaura, Bryan Scotneya, et al., "Smart City Architecture and its Applications based on IoT" | MLSC | Elsevier | 2015 |
| [S30] | Zhihong Yang, Yufeng Peng, et al., "Study and Application on the Architecture and Key Technologies for IOT" | - | IEEE | 2011 |
| [S31] | Miao Wu, Ting-lie Lu, et al. "Research on the architecture of Internet of things" | TMN | IEEE | 2010 |
| [S32] | Henrich C. Pohls, Vangelis Angelakis, "RERUM: Building a Reliable IoT upon Privacy- and Security- enabled Smart Objects" | RERUM | IEEE | 2014 |
| [S33] | ZAEI, "Reference Architecture Model Industry 4.0 (RAMI 4.0)" | RAMI | ZAEI | 2015 |
| [S34] | Pieter Ballon, Julia Glidden, "EPIC Platform and Technology Solution" | EPIC | EPIC | 2013 |
| [S35] | Kenji Tei, Levent G¨urgen, “ClouT : Cloud of Things for Empowering the Citizen Clout in Smart Cities” | ClouT | IEEE | 2014 |
| [S36] | Arup, “Solutions for Cities: An analysis of the feasibility studies from the Future Cities Demonstrator Programme” | TSB | Smart City Strategies A Global Review - ARUP | 2013 |
| [S37] | Liviu-Gabriel Cretu, Alexandru Ioan, “Smart Cities Design using Event-driven Paradigm and Semantic Web” | EdSC | Inforec Association | 2012 |
| [S38] | Luca Filipponi, Andrea Vitaletti, “Smart City: An Event Driven Architecture for Monitoring Public Spaces with Heterogeneous Sensors” | SOFIA | IEEE | 2010 |
| [S39] | Dan Puiu, Payam Barnaghi, et al., “CityPulse: Large Scale Data Analytics Framework for Smart Cities” | CityPulse | IEEE | 2016 |
| [S40] | ISO, “ISO/IEC 30182: Smart city concept model—Guidance for establishing a model for data interoperability” | SCCM | ISO (International Organization for Standardization) | 2017 |
| [S41] | Open Geospatial Consortium, “OGC Smart Cities Spatial Information Framework” | OGC | Open Geospatial Consortium | 2015 |
| [S42] | Roland Stühmer, Yiannis Verginadis, “PLAY: Semantics-Based Event Marketplace” | PLAY | Springer | 2013 |
| [S43] | M. Nitti, “IoT Architecture for a Sustainable Tourism Application in a Smart City Environment” | - | Hindawi | 2017 |
| [S44] | Carlos Costa, Maribel Yasmina Santos, “BASIS: A Big Data Architecture for Smart Cities” | BASIS | IEEE | 2016 |
| [S45] | BSI, “Smart city framework–Guide to establishing strategies for smart cities and communities” | BSI | British Standards Institution | 2014 |
| [S46] | S. J. Clement, D. W. McKee, “Service-Oriented Reference Architecture for Smart Cities” | SORASC | IEEE | 2017 |
| [S47] | Arundhati Bhowmick, Eduardo Francellino, et al., “IBM Intelligent Operations Center for Smarter Cities Administration Guide” | - | IBM | 2012 |
| [S48] | Andy Cox, Peter Parslow, et al., “ESPRESSO (systEmic Standardisation apPRoach to Empower Smart citieS and cOmmunities)” | ESPRESSO | ESPRESSO community | 2016 |
| [S49] | Arthur de M. Del Esposte , Fabio Kon, “InterSCity: A Scalable Microservice-based Open Source Platform for Smart Cities” | InterSCity | Scitepress digital library | 2017 |
| [S50] | Raffaele Giaffreda, “iCore: a cognitive management framework for the internet of things” | iCore | Springer | 2013 |
| [S51] | Andreas Kamilaris, Feng Gao, “Agri-IoT: A Semantic Framework for Internet of Things-enabled Smart Farming Applications” | Agri-IoT | IEEE | 2016 |
| [S52] | Yong Woo Lee, Seungwoo Rho, “U-City Portal for Smart Ubiquitous Middleware” | U-City | IEEE | 2010 |
| [S53] | Chayan Sarkar,Akshay Uttama Nambi S. N., “DIAT: A Scalable Distributed Architecture for IoT” | DIAT | IEEE | 2015 |
| [S54] | Gilles Privat, et al. “Towards a Shared Software Infrastructure for Smart Homes, Smart Buildings and Smart Cities” | SmartSantander | - | 2014 |
| [S55] | Tom Collins, “A Methodology for Building the IoT” | Collins | - | 2014 |
| [S56] | Frank Puhlmann, Dirk Slama, “An IoT Solution Methodology” | Ignite | - | Not stated |
| [S57] | C. Savaglio, "A Methodology for the Development of Autonomic and Cognitive Internet of Things Ecosystems" | ACOSO-Meth | - | 2017 |
| [S58] | G. Fortino, R. Gravina, et al., "A Methodology for Integrating Internet of Things Platforms" | INTER-METH | IEEE | 2018 |
| [S59] | Marcello A. Gómez Maureira, Daan Oldenhof, et al., "ThingSpeak–an API and Web Service for the Internet of Things" | ThingSpeak | World Wide Web | 2011 |
| [S60] | Venticinque Salvatore, Alba Amato, "A methodology for deployment of IoT application in fog" | BET | Springer | 2019 |
| [S61] | Amany Sarhan, "Cloud-based IoT Platform: Challenges and Applied Solutions" | Galliot | IGI Global | 2019 |
| [S62] | Alvaro Luis Bustamante , Miguel A. Patricio, "Thinger.io: An Open Source Platform for Deploying Data Fusion Applications in IoT Environments" | Thinger.io | Sensor | 2019 |
| [S63] | Fernando Terroso-Saenz, Aurora González, et al., "An open IoT platform for the management and analysis of energy data" | IoTEP | Elsevier | 2019 |

## Appendix B

An excerpt of the interview questions was used during the evaluation of the framework and the IoT-PA.

Part A: Evaluation criteria

1. Soundness

Do you think that the criteria defined in the IoT-PA are related and meant to be used for a platform assessment?

Please comment on any criterion that you think is not perceived as important for an assessment.

2. Generality

Do you think that the criteria defined in the IoT-PA are applicable for evaluating any IoT platform independent of its underlying technical implementation?

Please comment on any criterion that you think is too specific or domain specific, making it inapplicable to be used in assessment scenarios.

3. Simplicity

Do you think that the criteria defined in the IoT-PA are easy to understand and measure?

a. Please comment on any criterion whose definition is unclear.

b. Please comment on evaluation questions that are not clear to be used for examining a criterion.

4. Precision

Do you think the criteria are unambiguous and the leading questions enable to accurately measure a criterion?

Please comment on any evaluation questions that you think it is vague or coarse-grained and thus should be refined/decomposed into more questions to get an acceptable precision for an evaluation.

5. Minimum overlapping

Do you think that the proposed criteria are independent of each other and distinct?

6. Comprehensiveness

Do you think that the IoT-PA captures important and relevant criteria for incorporation into a scenario of IoT platform assessment?

Please comment on any missing criterion that should be captured by the IoT-PA.

Part B: Evaluation steps

1. Clarity of steps

Do you think the defined steps for assessing IoT platforms are clear and easy to understand?

2. Applicability

Do you think the IoT-PA enables users to compare and select a suitable IoT platform that needs their expected quality criteria?

Part C: Suggestions

In what ways do you think the IoT-PA creates value for users?

## Appendix C

Internal workshops to identify requirements the ROSE framework which provided input to the design phase of the DSR and our early version of the framework (with the kind permission from IoT Australia).

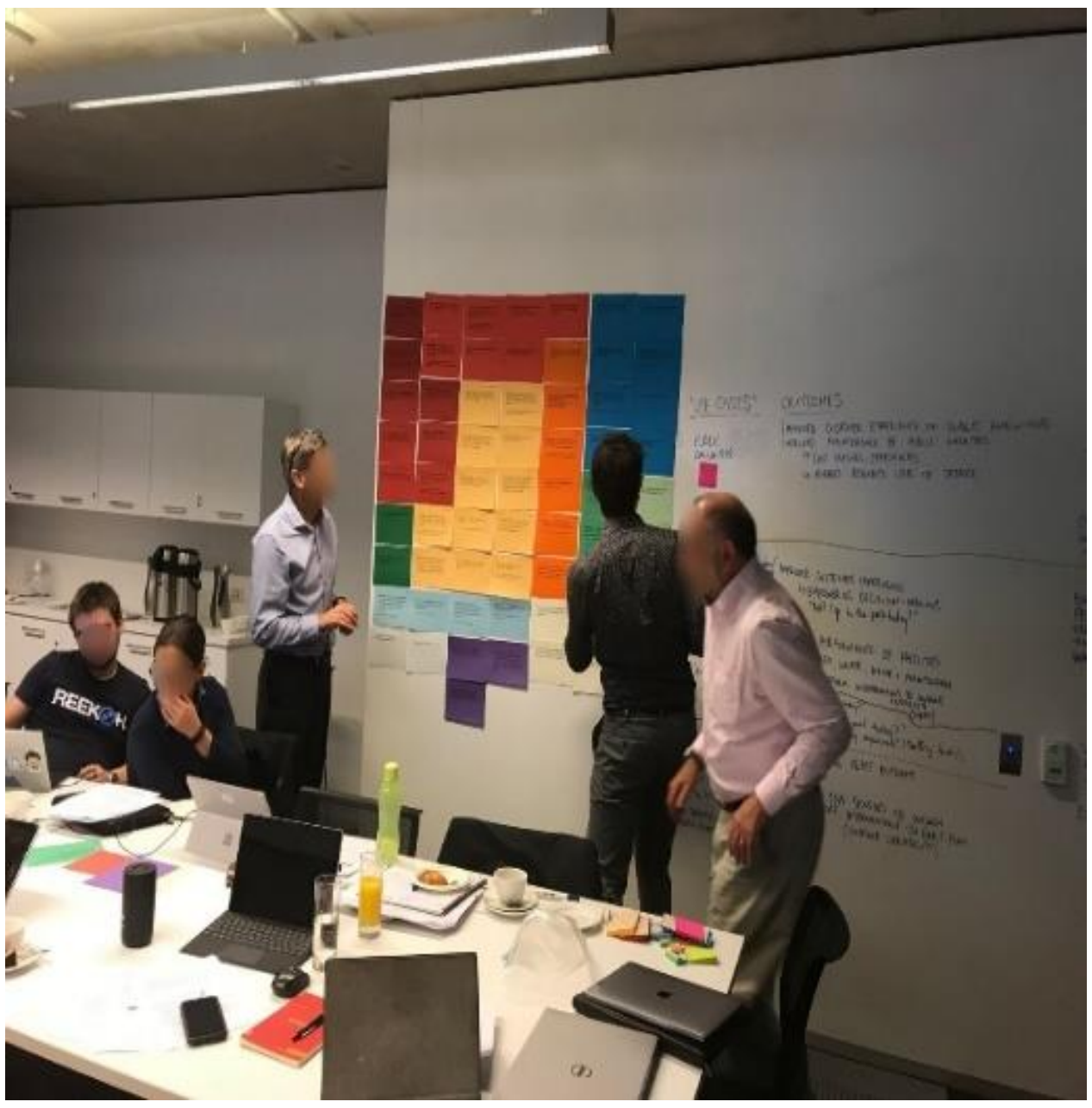

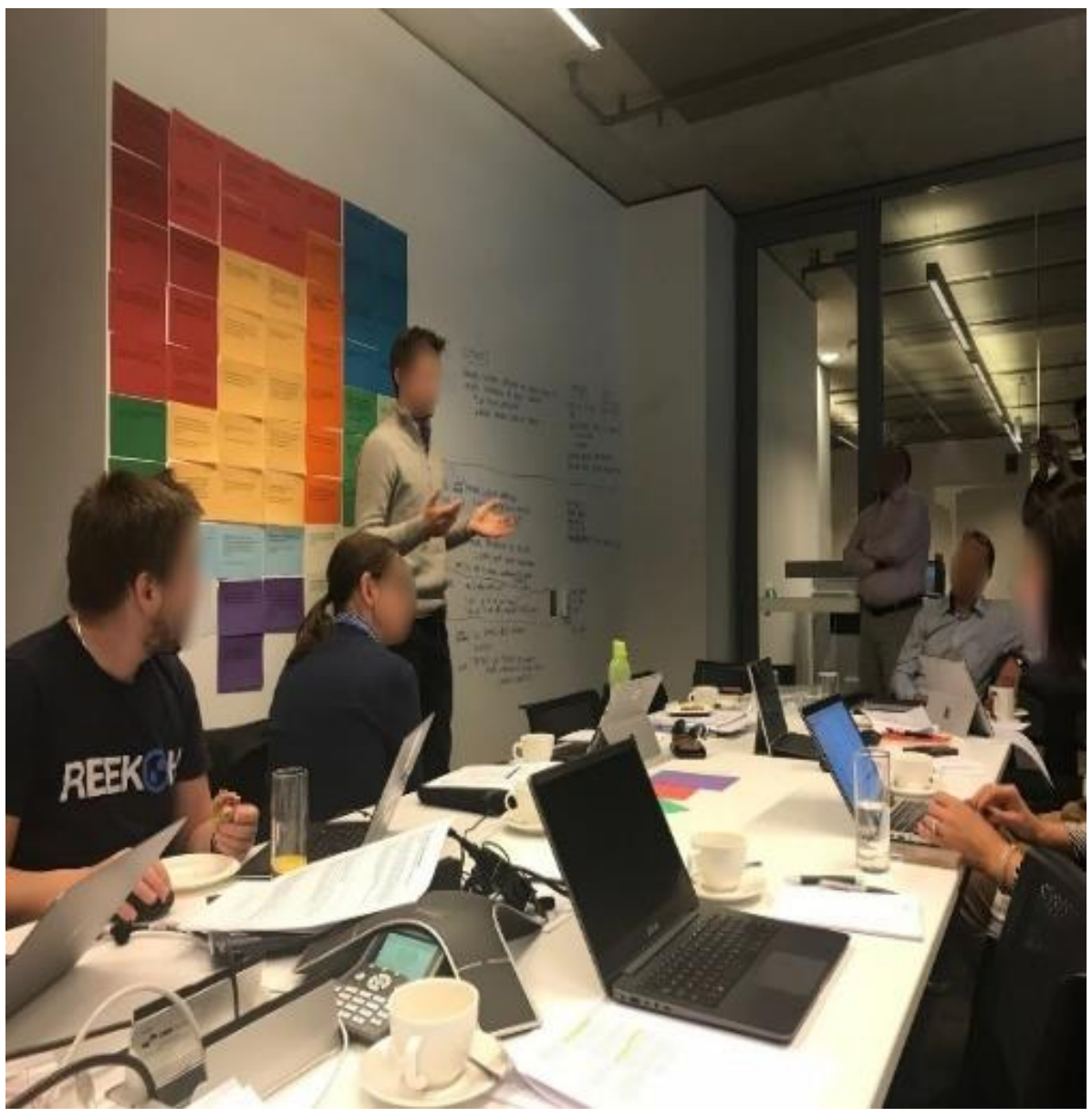

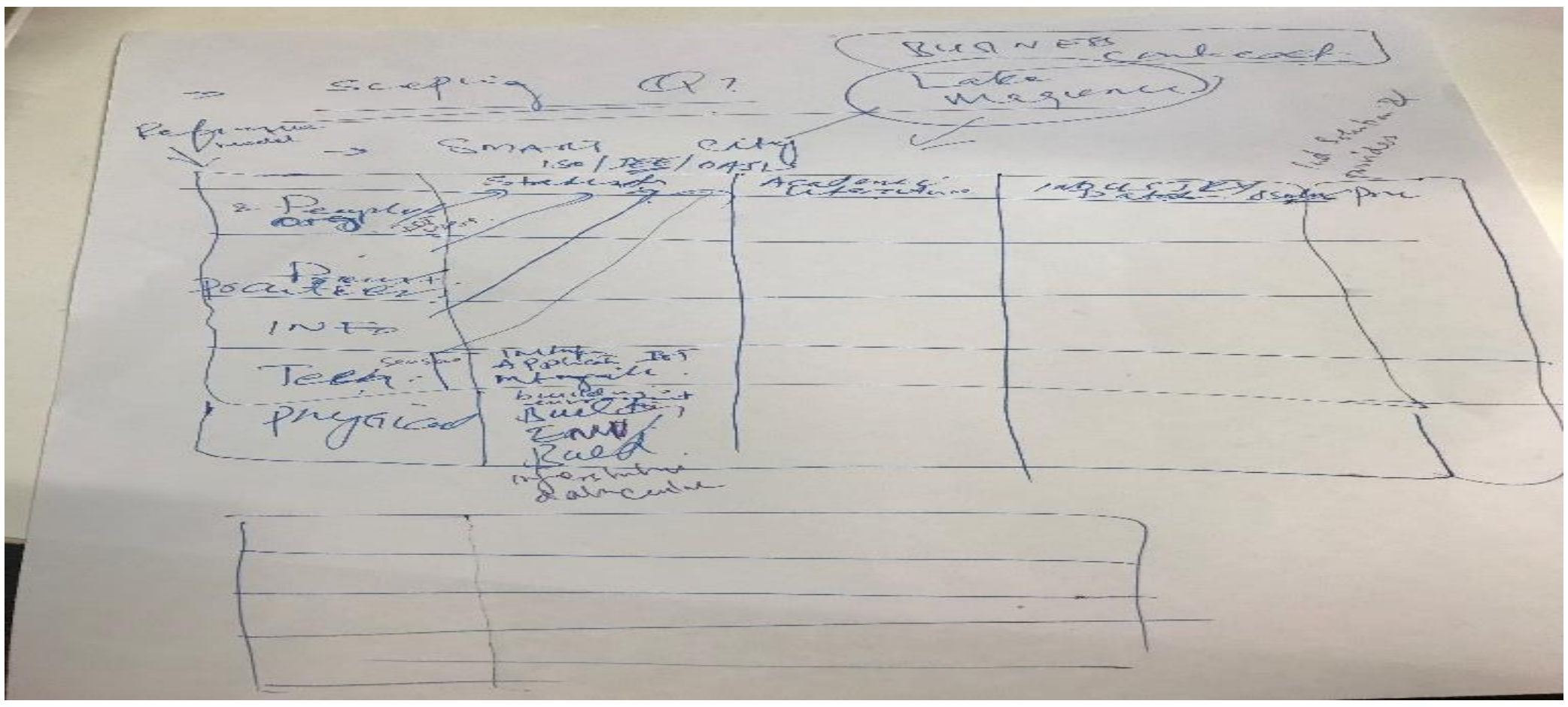

## References

1. Fahmideh, M.; Zowghi, D. An exploration of IoT platform development. *Inf. Syst.* **2020**, *87*, 101409.
2. Shin, D.-H. Conceptualizing and measuring quality of experience of the internet of things: Exploring how quality is perceived by users. *Inf. Manag.* 2017, *54*, 998–1011.
3. Li, L.; Rong, M.; Zhang, G. An Internet of Things QoE evaluation method based on multiple linear regression analysis. In Proceedings of the 2015 10th International Conference on Computer Science & Education (ICCSE), Cambridge, UK, 22–24 July 2015; IEEE: Piscataway, NJ, USA, 2015; pp. 925–928.

4. Bergman, J.; Olsson, T.; Johansson, I.; Rassmus-Gröhn, K. An exploratory study on how Internet of Things developing companies handle User Experience Requirements. In *International Working Conference on Requirements Engineering: Foundation for Software Quality*; Springer: Berlin/Heidelberg, Germany, 2018; pp. 20–36.
5. Henver, A.; March, S.T.; Park, J.; Ram, S. Design science in information systems research. *MIS Q.* 2004, *28*, 75–105.
6. Myers, M.D.; Venable, J.R. Management A set of ethical principles for design science research in information systems. *Inf. Manag.* 2014, *51*, 801–809.
7. Bass, L.; Clements, P.; Kazman, R. *Software Architecture in Practice*; Addison-Wesley Professional: Boston, MA, USA, 2003.
8. Komninos, N. *The Architecture of Intelligent Cities*; IET: London, UK, 2006.
9. Al-Hader, M.; Rodzi, A.; Sharif, A.R.; Ahmad, N. Smart city components architicture. In Proceedings of the Computational Intelligence, Modelling and Simulation, CSSim'09, International Conference on 2009, Brno, Czech Republic, 7–9 September 2009; IEEE: Piscataway, NJ, USA, 2019; pp. 93-97.
10. Dobrica, L.; Niemela, E. A survey on software architecture analysis methods. *Softw. Eng. IEEE Trans.* 2002, *28*, 638–653, doi:10.1109/tse.2002.1019479.
11. Bastidas, V.; Helfert, M.; Bezbradica, M. A Requirements Framework for the Design of Smart City Reference Architectures. In Proceedings of the 51st Hawaii International Conference on System Sciences, Hilton Waikoloa Village, HI, USA, 3–6 January 2018.
12. Santana, E.F.Z.; Chaves, A.P.; Gerosa, M.A.; Kon, F.; Milojicic, D.S. Software platforms for smart cities: Concepts, requirements, challenges, and a unified reference architecture. *ACM Comput. Surv. (CSUR)* 2017, *50*, 78.
13. Kitchenham, B.; Brereton, O.P.; Budgen, D.; Turner, M.; Bailey, J.; Linkman, S. Systematic literature reviews in software engineering—A systematic literature review. *Inf. Softw. Technol.* 2009, *51*, 7–15, doi:10.1016/j.infsof.2008.09.009.
14. Kitchenham, B.; Linkman, S.; Law, D. DESMET: A methodology for evaluating software engineering methods and tools. *Comput. Control Eng. J.* 1997, *8*, 120–126.
15. Taromirad, M.; Ramsin, R. An appraisal of existing evaluation frameworks for agile methodologies. In Proceedings of the 15th Annual IEEE International Conference and Workshop on the Engineering of Computer Based Systems (ecbs 2008), Belfast, UK, 31 March–4 April 2008; IEEE: Piscataway, NJ, USA, 2008; pp. 418–427.
16. Fahmideh, M. [Online: Auxiliary Review Material] A Comprehensive Framework for Analyzing IoT Platforms: A Smart City Industrial Experience. Available online: https://www.researchgate.net/publication/350579905_Auxiliary_material_-_a_list_of_IoT-specific_architecture_evaluation_criteria (accessed on April 2021).
17. Wohlin, C. Guidelines for snowballing in systematic literature studies and a replication in software engineering. In Proceedings of the 18th International Conference on Evaluation and Assessment in Software Engineering, London, UK, 13–14 May 2014; ACM: New York, NY, USA, 2014; p. 38.
18. Kakarontzas, G.; Anthopoulos, L.; Chatzakou, D.; Vakali, A. A conceptual enterprise architecture framework for smart cities: A survey based approach. In Proceedings of the e-Business (ICE-B), 2014 11th International Conference on 2014, Vienna, Austria, 28–30 August 2014; IEEE: Piscataway, NJ, USA, 2014; pp. 47–54.
19. da Silva, W.M.; Alvaro, A.; Tomas, G.H.; Afonso, R.A.; Dias, K.L.; Garcia, V.C. Smart cities software architectures: A survey. In Proceedings of the 28th Annual ACM Symposium on Applied Computing, Coimbra, Portugal, 18–22 March 2013; ACM: New York, NY, USA, 2013; pp. 1722–1727.
20. Al-Fuqaha, A.; Guizani, M.; Mohammadi, M.; Aledhari, M.; Ayyash, M. Internet of things: A survey on enabling technologies, protocols, and applications. *IEEE Commun. Surv. Tutor.* 2015, *17*, 2347–2376.
21. Triantaphyllou, E. Multi-criteria decision making methods. In *Multi-Criteria Decision Making Methods: A Comparative Study*; Springer: Berlin/Heidelberg, Germany, 2000; pp. 5–21.
22. Saaty, T.L. *The Analytic Hierarchy Process: Planning, Priority Setting, Resources Allocation*; M cGraw-Hill: New York, NY, USA, 1980.
23. van Looy, A.; de Backer, M.; Poels, G.; Snoeck, M. Choosing the right business process maturity model. *Inf. Manag.* 2013, *50*, 466–488.
24. Ngai, E. Selection of web sites for online advertising using the AHP. *Inf. Manag.* 2003, *40*, 233–242.
25. Liu, D.-R.; Shih, Y.-Y. Integrating AHP and data mining for product recommendation based on customer lifetime value. *Inf. Manag.* 2005, *42*, 387–400.
26. Benlian, A. Is traditional, open-source, or on-demand first choice? Developing an AHP-based framework for the comparison of different software models in office suites selection. *Eur. J. Inf. Syst.* 2011, *20*, 542–559.
27. Crawford, G. The geometric mean procedure for estimating the scale of a judgement matrix. *Math. Model.* 1987, *9*, 327–334.
28. Gill, A.Q.; Chew, E. Configuration information system architecture: Insights from applied action design research. *Inf. Manag.* 2019, *56*, 507–525.
29. Clements, P.; Kazman, R.; Klein, M. *Evaluating Software Architectures: Methods and Case Studies*; Addison-Wesley Reading: Boston, MA, USA, 2002.
30. Lassing, N.; Bengtsson, P.; van Vliet, H.; Bosch, J. Experiences with ALMA: Architecture-level modifiability analysis. *J. Syst. Softw.* 2002, *61*, 47–57.
31. Dolan, T.J. Architecture Assessment of Information-System Families. Ph. D Thesis, Eindhoven University of Technology, Department of Technology Management, Eindhoven, The Netherlands, 2002.

32. Ahmad, A.; Fahmideh, M.; Altamimi, A.B. Software Engineering for IoT-Driven Data Analytics Applications. *IEEE Access* 2021, *9*, 48197–48217.
33. Garg, S.K.; Versteeg, S.; Buyya, R. A framework for ranking of cloud computing services. *Future Gener. Comput. Syst.* 2013, *29*, 1012–1023.
34. Khajeh-Hosseini, A.; Sommerville, I.; Bogaerts, J.; Teregowda, P. Decision support tools for cloud migration in the enterprise. In Proceedings of the Cloud Computing (CLOUD), 2011 IEEE International Conference on 2011, Washington, DC, USA, 4–9 July 2011; IEEE: Piscataway, NJ, USA, 2011; pp. 541–548.
35. Guth, J.; Breitenbücher, U.; Falkenthal, M.; Fremantle, P.; Kopp, O.; Leymann, F.; Reinfurt, L. A Detailed Analysis of IoT Platform Architectures: Concepts, Similarities, and Differences. In *Internet of Everything*; Springer: Berlin/Heidelberg, Germany, 2018; pp. 81–101.
36. Guth, J.; Breitenbücher, U.; Falkenthal, M.; Leymann, F.; Reinfurt, L. Comparison of IoT platform architectures: A field study based on a reference architecture. In Proceedings of the Cloudification of the Internet of Things (CIoT), Paris, France, 23–25 November 2016; IEEE: Piscataway, NJ, USA, 2016; pp. 1–6.
37. Kyriazopoulou, C. Smart city technologies and architectures: A literature review. In Proceedings of the Smart Cities and Green ICT Systems (SMARTGREENS), 2015 International Conference on 2015, Lisbon, Portugal, 20–22 May 2015; IEEE: Piscataway, NJ, USA, 2015; pp. 1–12.
38. Choi, J.; Kim, S. An AHP Approach toward Evaluating IoT Business Ecosystem in Korea. In Proceedings of the 29th European Regional Conference of the International Telecommunications Society (ITS), Trento, Italy, 1–4 August 2018.
39. Huang, Y.-L.; Sun, W.-L. An AHP-Based Risk Assessment for an Industrial IoT Cloud. In Proceedings of the 2018 IEEE International Conference on Software Quality, Reliability and Security Companion (QRS-C), Lisbon, Portugal, 16–20 July 2018; IEEE: Piscataway, NJ, USA, 2018; pp. 637–638.
40. Mahdi, F.; Didar, Z. IoT Architectures: An analytical review. In Proceedings of the 9th IEEE Annual Information Technology, Electronics and Mobile Communication Conference, Vancouver, BC, Canada, 1–3 November 2018; IEEE: Piscataway, NJ, USA, 2018.
41. Kondratenko, Y.; Kondratenko, G.; Sidenko, I. Multi-criteria decision making for selecting a rational IoT platform. In Proceedings of the 2018 IEEE 9th International Conference on Dependable Systems, Services and Technologies (DESSERT), Kyiv, Ukraine, 24–27 May 2018; IEEE: Piscataway, NJ, USA, 2018; pp. 147–152.
42. Contreras-Masse, R.; Ochoa-Zezzatti, A.G. Selection of IoT Platform with Multi-Criteria Analysis: Defining Criteria and Experts to Interview. *Res. Comput. Sci.* 2019, *148*, 9–19.
43. Silva, E.M.; Jardim-Goncalves, R. IoT Ecosystems Design: A Multimethod, Multicriteria Assessment Methodology. *IEEE Internet Things J.* 2020, *7*, 10150–10159.